\documentclass[conference,10pt]{IEEEtran}

\usepackage{listings}
\lstdefinestyle{promptstyle}{
    basicstyle=\ttfamily\small,
    frame=single,
    rulecolor=\color{black},
    breaklines=true,
    numbers=none,
    tabsize=4,
    keywordstyle=\color{blue},
    commentstyle=\color{green!50!black},
    stringstyle=\color{red}
}

\usepackage{longtable}
\usepackage{booktabs}   
\usepackage{subcaption} 

\usepackage{svg}
\usepackage{booktabs}
\usepackage{tabularx}
\usepackage{tikz}
\usepackage{comment}

\usepackage{array}
\usepackage{multirow}
\usepackage{multicol}
\usepackage{hhline}
\usepackage[export]{adjustbox}
\usepackage{graphicx}
\usepackage{makecell}

\usepackage{threeparttable}
\usepackage{url}

\usepackage[ruled,vlined]{algorithm2e}
\usepackage{setspace}
\usepackage{algorithmic}
\usepackage[skip=0pt,font=small]{subfig}
\usepackage[skip=0pt,font=small]{caption}
\usepackage{todonotes}
\usepackage{enumitem}

\usepackage{tikz}
\usetikzlibrary{calc}

\usepackage{amsmath}
\usepackage{amssymb}
\usepackage{caption}

\DeclareRobustCommand*\circled[1]{\tikz[black,baseline=(char.base)]{
            \node[shape=circle,fill,inner sep=1.3pt] (char) {\textcolor{white}{#1}};}}

\definecolor{myblue}{RGB}{91,155,213}
\definecolor{mydark}{RGB}{0,0,0}

\newcommand{\chen}[1] {{\color{green!80!black}{#1}}}
\newcommand{\wahib}[1] {{\color{blue}{#1}}}

\newcommand{\method}{\texttt{ROVAI}}

\begin{document}


\title{{\fontsize{23pt}{26pt}\selectfont Paradigm Shift in Infrastructure Inspection Technology}
}

\author{
  \IEEEauthorblockN{Du Wu\textsuperscript{1,2, *}, Enzhi Zhang\textsuperscript{3, *}, 
  Isaac Lyngaas\textsuperscript{4}, Xiao Wang\textsuperscript{4}, Amir Ziabari\textsuperscript{4},
  Tao Luo\textsuperscript{5}, Peng Chen\textsuperscript{1}, Kento Sato\textsuperscript{1}, \\ 
  Fumiyoshi Shoji\textsuperscript{1}, Takaki Hatsui\textsuperscript{6}, Kentaro Uesugi\textsuperscript{6}, Akira Seo\textsuperscript{7}, Yasuhito Sakai\textsuperscript{8}, \\
  Toshio Endo\textsuperscript{2}, Tetsuya Ishikawa\textsuperscript{6}, Satoshi Matsuoka\textsuperscript{1, $\dagger$}, Mohamed Wahib\textsuperscript{1, $\dagger$}}
  \IEEEauthorblockA{
  \textit{\textsuperscript{1} RIKEN Center for Computational Science, Japan}, 
  \textit{\textsuperscript{2} Institute Science Tokyo, Japan}, 
 \textit{\textsuperscript{3} Hokkaido University, Japan}\\
 \textit{\textsuperscript{4} Oak Ridge National Laboratory, USA}, 
 \textit{\textsuperscript{5} A*STAR, Singapore},
 \textit{\textsuperscript{6} JASRI, SPring-8, Japan}\\
 \textit{\textsuperscript{7} Kyoto University, Japan}, 
 \textit{\textsuperscript{8} Hanshin Expressway Public Corporation, Japan}\\ 
 } 
}
\maketitle
\thispagestyle{plain}
\pagestyle{plain}
\let\thefootnote\relax\footnotetext{ *Co-first author: \emph{Du Wu} and \emph{Enzhi Zhang} are equal contributors}
\let\thefootnote\relax\footnotetext{$\dagger$Corresponding author: \emph{Satoshi Matsuoka} (matsu@acm.org), \emph{Mohamed Wahib} (mohamed.attia@riken.jp)}
\begin{abstract}
Effective road infrastructure management is crucial for modern society. Traditional manual inspection techniques remain constrained by cost, efficiency, and scalability. By leveraging the computational power of world-leading flagship supercomputers, Fugaku and Frontier, and one of the world's leading synchrotron facilities, SPring-8, our framework, \method{}, enables scalable and high-throughput processing of massive 3D tomographic datasets.
Our approach overcomes key challenges to: 
a) scale the simultaneous reconstruction of high-resolution X-ray computed tomography images, 
b) introduce novel solutions for handling high memory requirements and the lack of labeled training data when training vision transformer models on Frontier, and 
c) develop an end-to-end pipeline that seamlessly integrates imaging and AI analytics at the full Fugaku system scale. 
\method{} sets a new standard for intelligent, data-driven infrastructure management in real-world scenarios, enabling automated defect detection, material composition analysis, and lifespan prediction of highway pavements. With \method{}, we reconstruct a total of 46 specimens into 8{,}192-resolution 3D images, followed by AI segmentation inference in a total of 2{,}762 seconds (on average, 1 full specimen per 60 seconds), achieving approximately 95\% Dice score image segmentation accuracy. This is achieved on the full scale of Fugaku (152{,}064 compute nodes), with a sustained storage I/O bandwidth reaching 60~GB/s, corresponding to a system-level I/O utilization rate of 40\%.

\end{abstract}
\section{Justification for ACM Gordon Bell Prize}
ROVAI transforms and sets a new standard for intelligent, data-driven infrastructure management in real-world scenarios. We reconstruct a total of 46  Three-Dimensional (3D) X-ray Computed
Tomography (XCT) specimens into 8,192 resolution 3D images, fused in a single pipeline with AI segmentation inference in a total of 2,762 seconds (1 full 8TB specimen per 60 seconds): achieving $\sim$95\% image segmentation accuracy. Achieving end-to-end sustained storage bandwidth of 40\% of full scale of Fugaku supercomputer (152,064 compute nodes).


\begin{figure}[t]
  \begin{center}
    \includegraphics[clip,width=0.49\textwidth]{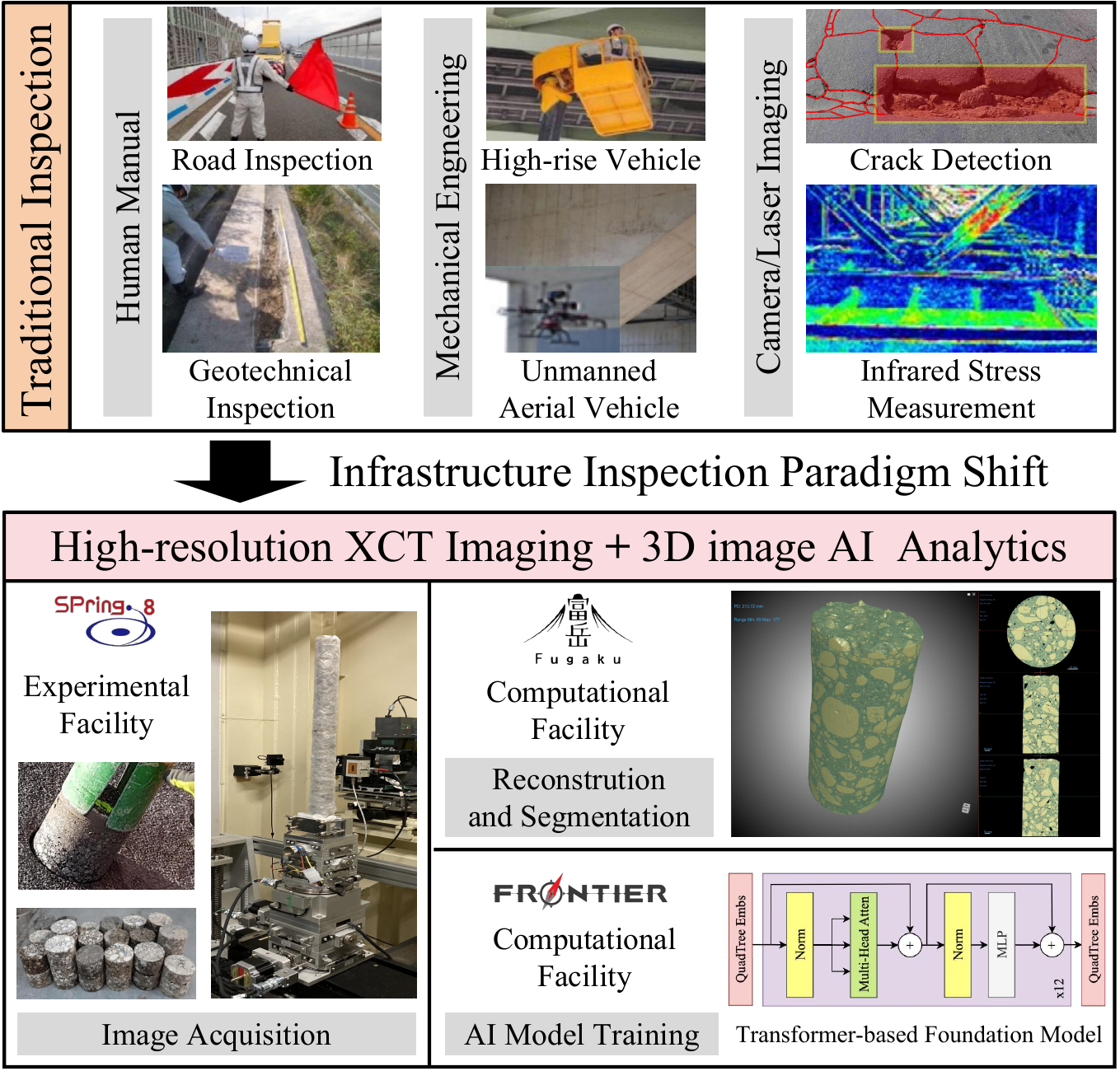}
    \caption{\small{
    This work transforms infrastructure inspection towards a technology based on high-performance imaging and AI analytics. We use three flagship facilities in a coordinated manner:
    \textbf{a)} RIKEN SPring-8 synchrotron (experimental facility): providing high-quality XCT scans of infrastructure specimens in a fast turnaround, while managing high-volume data transfer to other facilities, \textbf{b)} RIKEN Fugaku supercomputer (computational facility): a large-scale system with 158,976 compute nodes, offering a balanced design for data movement (network and storage) and capable of processing dozens of high-resolution volume reconstructions fused with AI analytics, and \textbf{c)} ORNL Frontier supercomputer (computational facility): a powerful system equipped with 37,888 GPUs, enabling more effective large-scale and efficient pre-training of foundational vision models.
    }}
    \label{fig:intro}
  \end{center}
\end{figure}
\section{Performance Attributes}
\begin{center}
\resizebox{\linewidth}{!}{
\renewcommand{\arraystretch}{0.75}
\begin{tabular}{|c|c|}
\hline
Attributes & Contents \\
\hline
 Category & \begin{tabular}{@{}c@{}}{\em Time-to-solution}, {\em Scalability}, {\em Peak Perf.}\\ \end{tabular} \\
\hline
Type of method & \begin{tabular}{@{}c@{}}{\em Computed Tomography Fused}\\ {\em w/ Vision Transformer Model}\end{tabular}

\\
\hline
Results reported on & {\em Whole application including I/O}\\
\hline
Precision reported & {\em FP32 
(single precision)}\\
\hline
System scale & {\em Measured on Full-Scale System} \\
\hline
Measurement mechanism & \begin{tabular}{@{}c@{}}{\em Timer}, {\em FLOP counts}, {\em Job reports}\end{tabular} \\
\hline
\end{tabular}}
\label{Attributes}
\end{center}

\section{Overview of the Problem}\label{sec:introduction}
Road infrastructure forms the backbone of modern society. However, managing and maintaining infrastructure presents significant challenges for governments worldwide, with road maintenance alone costing thousands of billions of dollars annually~\cite{glaeser2021economic}.
Traditional inspection approaches that rely on mechanical instruments incur high costs and require significant time~\cite{barriera2021assessing}.
Camera and laser imaging technologies are useful for quickly screening visible surface cracks and depressions, yet they fall short in evaluating subsurface conditions~\cite{yao2023advanced}.

Advances in two particular technologies provide an opportunity to transform established methods for inspecting road infrastructure: XCT (X-ray Computed Tomography) and AI-based analytics driven by ViTs (Vision Transformers)~\cite{dosovitskiy2020image}. XCT scans of rock-based structural materials~\cite{kim2017investigation,zhao20213d} under various conditions, including manufacturing and usage environments, allow for identifying characteristic structural changes during deformation and failure~\cite{taheri2023review}. 
Vision Transformers such as SAM 2~\cite{ravi2024sam}, nnUNet~\cite{isensee2021nnu}, Swin-Unet~\cite{cao2022swin}, and TransUNet~\cite{DBLP:journals/corr/abs-2102-04306} are becoming increasingly effective at delivering high-quality segmentation for both scientific and engineering 2D and 3D images~\cite{zhang2024adaptive}. Moreover, these models can be fine-tuned for downstream tasks such as classifying and analyzing the distribution, size, and spacing of segmented objects.

In theory, advances in XCT imaging and vision AI analytics could together provide a foundation for analyzing material composition, aggregate cracks, deformation, strain, internal voids, connectivity with external environments, and humidity conditions at both molecular and structural levels. These capabilities, if enabled by imaging and advanced AI analytics, offer the potential to replace traditional methods, assess the current state of materials, and predict the lifespan of road infrastructure. However, several key challenges still hinder the realization of this qualitative technological transition.

\circled{1} XCT resolution has surged from $2K^2$ to $64K^2$\cite{dudipala2024halide}, increasing memory and compute demands by 1,024$\times$ and 4,096$\times$, respectively. Reconstructing $8K\times8K$ and $16K\times16K$ images is essential for resolving micro-structural features, yet traditional methods—focused on single-image optimization~\cite{bicer2017trace,wang2017massively,hidayetouglu2019memxct,hidayetouglu2020petascale,chen2019ifdk,chen2021scalable, wu2024real} 
struggle with parallelism at scale. \method{} enables high performance and throughput by reconstructing dozens of images simultaneously at full system scale. It also addresses the ballooning data sizes from modern synchrotron scans, which can exceed terabytes (e.g.~\cite{hidayetouglu2019memxct}), through scalable end-to-end pipelines.



\circled{2} 3D segmentation is the key enabler for analytics, allowing lightweight models to handle downstream tasks like identifying material type, size, distribution, and mixing conditions. However, high-res XCT images lack ground truth labels, making manual annotation impractical (e.g., 28,800 slices for an $8K\times8K\times28,800$ volume). To address this, \method{} pre-trains a 3D segmentation foundation model using a two-stage process: a) self-supervised learning with Masked Autoencoders (MAEs)~\cite{he2022masked}, and b a novel mask generation technique based on simulating microstructure and density features to produce accurate labels with minimal manual input.

\circled{3} The high memory requirements of state-of-the-art vision transformer models like SAM 2~\cite{ravi2024sam} and Swin~\cite{liu2021swin} present another difficulty, as these models struggle to process high-resolution images. \method{} addresses this limitation by taking inspiration from adaptive patching techniques~\cite{zhang2024adaptive} to use symmetric adaptive patching, which eliminates the need for memory-expensive decoders used in ViTs, hence reducing the required computation and memory by orders of magnitude while preserving model accuracy.

\circled{4} PFS I/O bottlenecks arise from traditional workflows that separate image reconstruction and AI analysis, causing repeated data movement. \method{} eliminates this by fusing both into a single memory-resident pipeline, enabling continuous X-ray streaming and fast processing. Since XCT reconstruction and AI inference use different data distribution strategies, 3D parallelism for reconstruction vs. model-optimized partitioning, \method{} adapts the inference scheme to the reconstructed sub-volumes in memory, avoiding costly load/store operations of tens of terabytes.

Ensuring \method{} produces high-quality output is crucial for various downstream applications, including defect detection and quality assessment, supporting automated decision-making and large-scale infrastructure assessments. By addressing these challenges, \method{} emerges as a transformative AI-driven solution for industrial imaging through three key breakthroughs:
\begin{itemize}[leftmargin=2.5mm]
\item \textbf{Scalable high-performance and high-throughput XCT:} A high-performance parallel-beam XCT pipeline was optimized for supercomputers, enabling rapid processing of $8K^2$–$16K^2$ images totaling hundreds of terabytes on Fugaku. We achieved 3D reconstruction of dozens of $8K^2$ specimens in under a minute, reaching 37.1\% (28 PFLOPS) of peak performance on 12,288 nodes. The pipeline reconstructed asphalt and concrete specimens from western Japan highways scanned at SPring-8, producing high-quality data for training a foundation AI model.
\item \textbf{State-of-the-Art ViT High-resolution 3D Segmentation Accuracy:} We pre-trained and fine-tuned a foundation vision transformer for 3D segmentation of high-resolution images of concrete and pavement using thousands of GPUs on the Frontier supercomputer. Due to the impracticality of labeling 8K and 16K resolution 3D images, we first used unsupervised training and developed a novel method to simulate ground truth images with corresponding masks. We also introduced a symmetric patch-based approach that eliminates the need for a decoder. Our method achieves ~95\% segmentation accuracy on 8K images, outperforming the state-of-the-art SAM 2~\cite{ravi2024sam} model (86\%). We also demonstrate how the segmented 3D images can be used for downstream tasks such as classification and analysis of the distribution, size, and spacing of segmented objects.
\item \textbf{Efficient Fusion of XCT and AI Inference:} We developed an end-to-end framework, \method{}, which combines the image reconstruction pipeline with inference from the pre-trained AI model to enable processing of new infrastructure specimens expected to be collected in the coming years. \method{} seamlessly fuses reconstruction and inference, memoized pipelining, minimizing data transfer bottlenecks (storage I/O), and improving overall computing performance. On a full-scale system such as Fugaku, reconstruction of dozens of 8K resolution 3D images, combined with AI-based 3D segmentation, can be completed in under a minute per image. Our approach has been validated on real-world national infrastructure specimens, demonstrating both practical scalability and effectiveness.

\end{itemize}

\begin{figure*}[t]
  \begin{center}
    \includegraphics[clip,width=\textwidth]{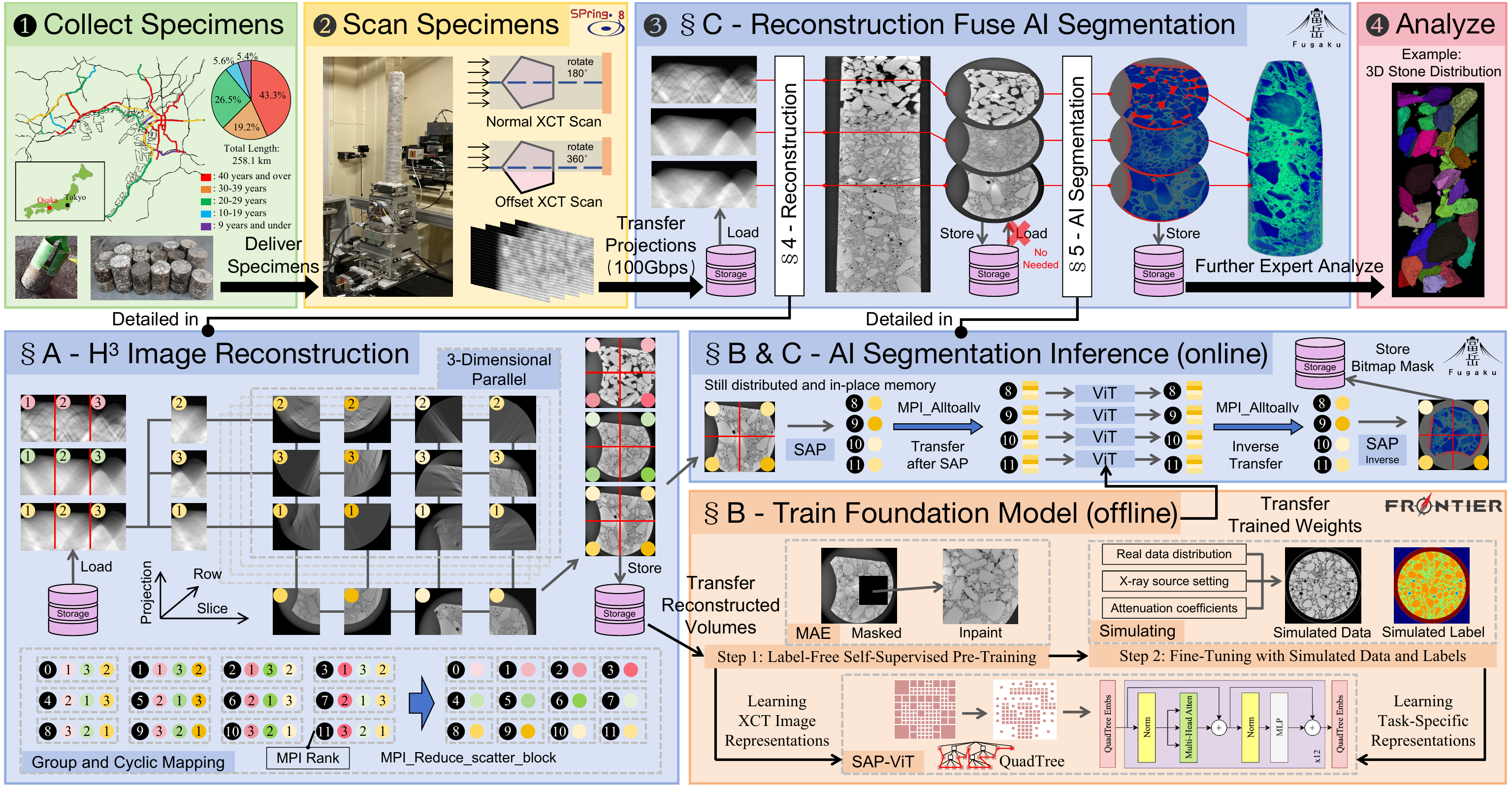}    \caption{\small{Overview of \method{}: an end-to-end high-performance imaging and AI-driven analysis framework for road infrastructure inspection.}}
    \label{fig:overview}
  \end{center}
\end{figure*}
\section{Current State of the Art}\label{sec:background}
\subsection{Classical Road Inspection Methods} 
Conventional road inspection methods (Fig.~\ref{fig:intro}) face cost, efficiency, and scalability challenges. Manual inspection is straightforward but labor-intensive and prone to inconsistency. Mechanical techniques, such as deflectometers and ground-penetrating radar, provide subsurface information but are typically slow and limited in capability. Camera and laser-based systems enable efficient detection of surface defects, yet are unable to detect subsurface damage.

\subsection{Parallel-beam XCT Image Reconstruction}
Filtered Back Projection (FBP)~\cite{kak2001principles} is a widely used solver for reconstructing images in parallel-beam XCT. It consists of two main steps: filtering and BP (Back Projection).
First, projections collected from different angles are filtered to enhance important details and reduce blurring. Next, the filtered projections are combined to form the final image through the BP operation to be mapped back onto the image space.

\subsection{Vision AI model and Image Segmentation} 
Vision Transformers (ViT)~\cite{dosovitskiy2020image} model long-range dependencies by processing images as sequences of fixed-size patches (e.g. $16\times16$). This enables the capturing of spatial relationships, making ViT well-suited for image segmentation. However, standard ViT models struggle with high-resolution images due to computational and memory limits. To overcome this, recent research has introduced adaptive patching, which dynamically adjusts patch sizes based on spatial complexity. This approach~\cite{zhang2024adaptive} improves segmentation accuracy by reducing redundant computations in simple areas while preserving details in complex regions.



\subsection{Overview of \method}
Fig.~\ref{fig:overview} presents an overview of \method. Dozens of specimens have been collected, and collection will continue from the national highway network of the Kansai urban region of western Japan. This area is home to over 20 million people, with major cities including Osaka, Kyoto, and Kobe. The highway network spans more than 250 km and carries approximately 700,000 vehicles daily. Since over 2/3 of the network is more than 30 years old, it is essential to develop effective technologies to identify road segments that are prone to critical failures. These include delamination, cohesive failure, insufficient asphalt coverage around aggregates, and cracking of the aggregates themselves. The specimens are transported to RIKEN SPring-8, a flagship synchrotron radiation facility, where they are scanned in high-resolution XCT. The scans were then moved to the RIKEN Fugaku supercomputer facility, where a high-throughput image reconstruction pipeline we developed was used to reconstruct the 3D images of dozens of specimens with tens of thousands of compute nodes. While Fugaku is a capable supercomputer, its processors lack matrix processing units and support for lower precision. To that end, effective training of a foundation model was done on a supercomputer more suitable for the task. The 3D images reconstructed on Fugaku were then moved to the ORNL Frontier supercomputer to train a Vision Transformer foundation model for 3D volumetric segmentation on thousands of GPUs. The foundation model includes key innovations, namely a simulator for generating masks and symmetric adaptive patching. Next, the model was fine-tuned, and the model weights were moved to the Fugaku supercomputer. Finally, we conducted full-scale Fugaku runs to demonstrate the ability to simultaneously reconstruct dozens of specimens followed by AI inference in a single fused end-to-end pipeline, from loading the scans from the PFS I/O to writing back the 3D segmented images to storage.



\section{Innovation Realized}

\subsection{H\textsuperscript{3}: High-throughput, High-perf., High-res. Imaging}\label{sec:H3}



\begin{figure}[t]
\begin{center}

\begin{minipage}{0.485\textwidth}
    \captionof{table}{\small{ 
        Parameters used in \method{}. The upper section lists acquisition-specific parameters, the lower section defines parallelization settings.}}        
    \resizebox{0.98\linewidth}{!}{
        \begin{tabular}{ |c|l| }
        \hline
        \bf{Symbol} & \bf{Description} \\
        \hline
        $N_{specimen}$  & Number of specimens \\
        $N_p^i$ & Number of projection angles in $i^{\text{th}}$ specimen \\
        $N_c^i$ & Detector channels of the 2D projection in $i^{\text{th}}$ specimen  \\
        $N_r^i$ & Detector rows of the 2D projection in $i^{\text{th}}$ specimen \\
        $N_x^i, N_y^i, N_z^i$ & Dimensions of 3D volume in the X, Y, and Z in $i^{\text{th}}$ volume \\
        \hline
        $P_{row}$ & Number of row parallel MPI ranks \\
        $P_{proj}$ & Number of projection parallel MPI ranks \\
        $P_{slice}$ & Number of slice parallel MPI ranks \\
        $\mathbb{G}$ & Set of group row partitions\\
        \hline
        \end{tabular}
        }
    \label{tab:param}
\end{minipage}

\vskip 5pt

\begin{minipage}{0.485\textwidth}
    \includegraphics[clip,width=0.98\textwidth]{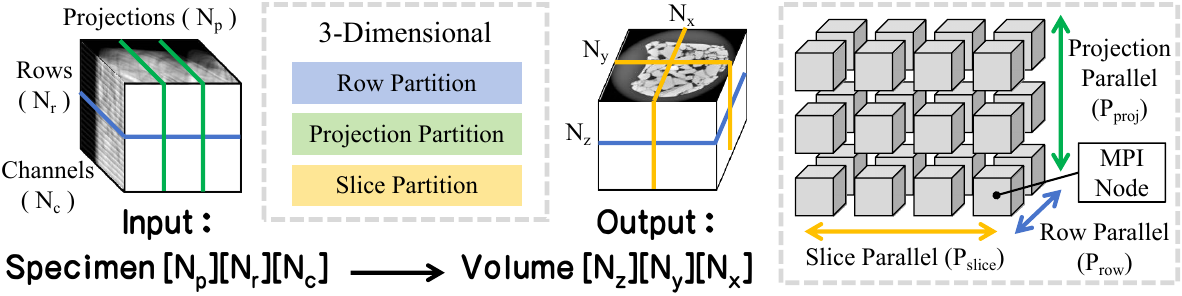}

  \caption{\small{Illustration of how a single specimen image is reconstructed using 3-dimensional partitioning and parallelism.}}
    \label{fig:3D-partitioning}
\end{minipage}

\end{center}
\end{figure}

\method{} leverages the massive parallelism of the Fugaku supercomputer to process high-resolution X-ray images acquired by XCT scanners at the Spring-8 synchrotron facility.
This section outlines the design of a distributed reconstruction pipeline that efficiently processes multiple volumes in parallel, balancing computation, communication, and I/O to minimize overall end-to-end runtime.
Table~\ref{tab:param} summarizes the key parameters used in configuring \method{}.

\begin{figure*}[t]
  \begin{center}
\includegraphics[clip,width=\textwidth]{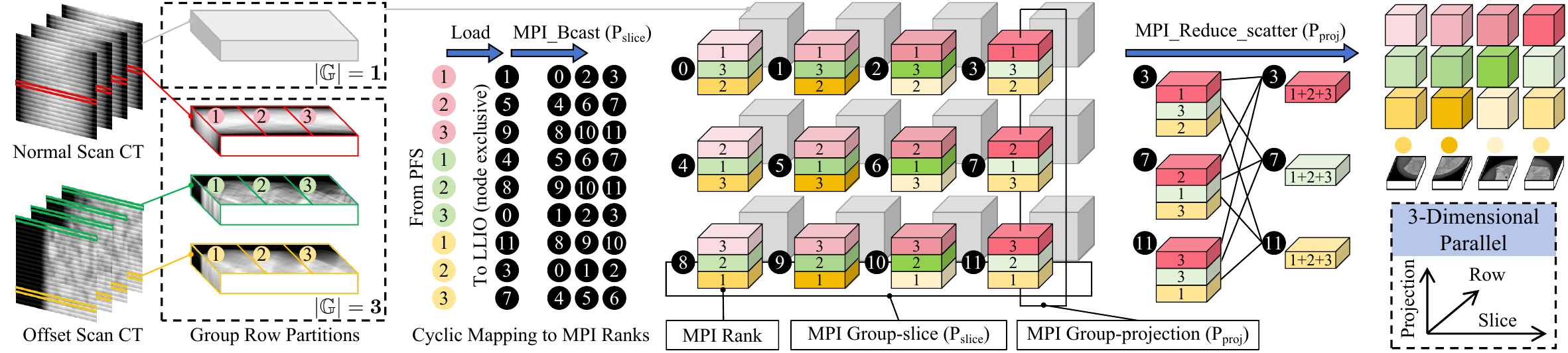}
    \caption{\small{Proposed H\textsuperscript{3} imaging algorithm in \method{}: Group row partitions and cyclic mapping to MPI ranks.}}
    \label{fig:partitioning}
  \end{center}
\end{figure*}

\textbf{a) 3D Partitioning for Distributed Computing.}\label{sec:parallel}
To enable fine-grained parallelism, Fig.~\ref{fig:partitioning} introduces a three-dimensional partitioning and parallelization strategy we use in our framework. As shown in Fig.~\ref{fig:3D-partitioning}, the image reconstruction, transforming projections to volume data, is divided along (input) scan-rows, (input) projections, and (output) 2D slices from the 3D volume, with each dimension supporting parallel computation.

\underline{Scan-rows Partition.}
In parallel beam XCT, each detector row directly maps to a distinct volume slice. Because the projections captured by each row are independent, reconstruction can be performed separately for each slice without cross-dependencies. As illustrated in Fig.~\ref{fig:3D-partitioning}, we partition the data in the $N_r$ dimension to enable distributed computation.
This makes the row dimension naturally partitionable, allowing straightforward parallelization across both the specimen and volume.
\underline{Projections Partition.}
Projections, composed of a series of scan-rows, are acquired at evenly spaced rotational angles. As illustrated in Fig.~\ref{fig:3D-partitioning}, we partition the data along the $N_p$ dimension to enable parallel computation. 
Each 2D projection contributes to updating the entire volume. The final reconstructed volume is obtained by summing multiple partial volumes, each generated from a subset of projections. Therefore, after projection partitioning and parallel processing, a reduction operation \textit{MPI\_Reduce} is required to generate the final volume.
\underline{Slices Partition.}
Since each voxel in the output volume is computed independently, thus allowing for partitioning of 2D slices from the 3D volume.
Since adjacent voxels exhibit improved data locality in both projections and volumes during distributed FBP computation, we partition in both the $N_x$ and $N_y$ dimensions to enhance data locality and achieve fine-grained parallelism. The three dimensions of parallelism work together to optimize both computation and communication in image reconstruction. In Section~\ref{sec:balancing}-b, we will address the load imbalance introduced by parallelism.
\begin{figure}[t]
  \begin{center}
    \includegraphics[clip,width=0.475\textwidth]{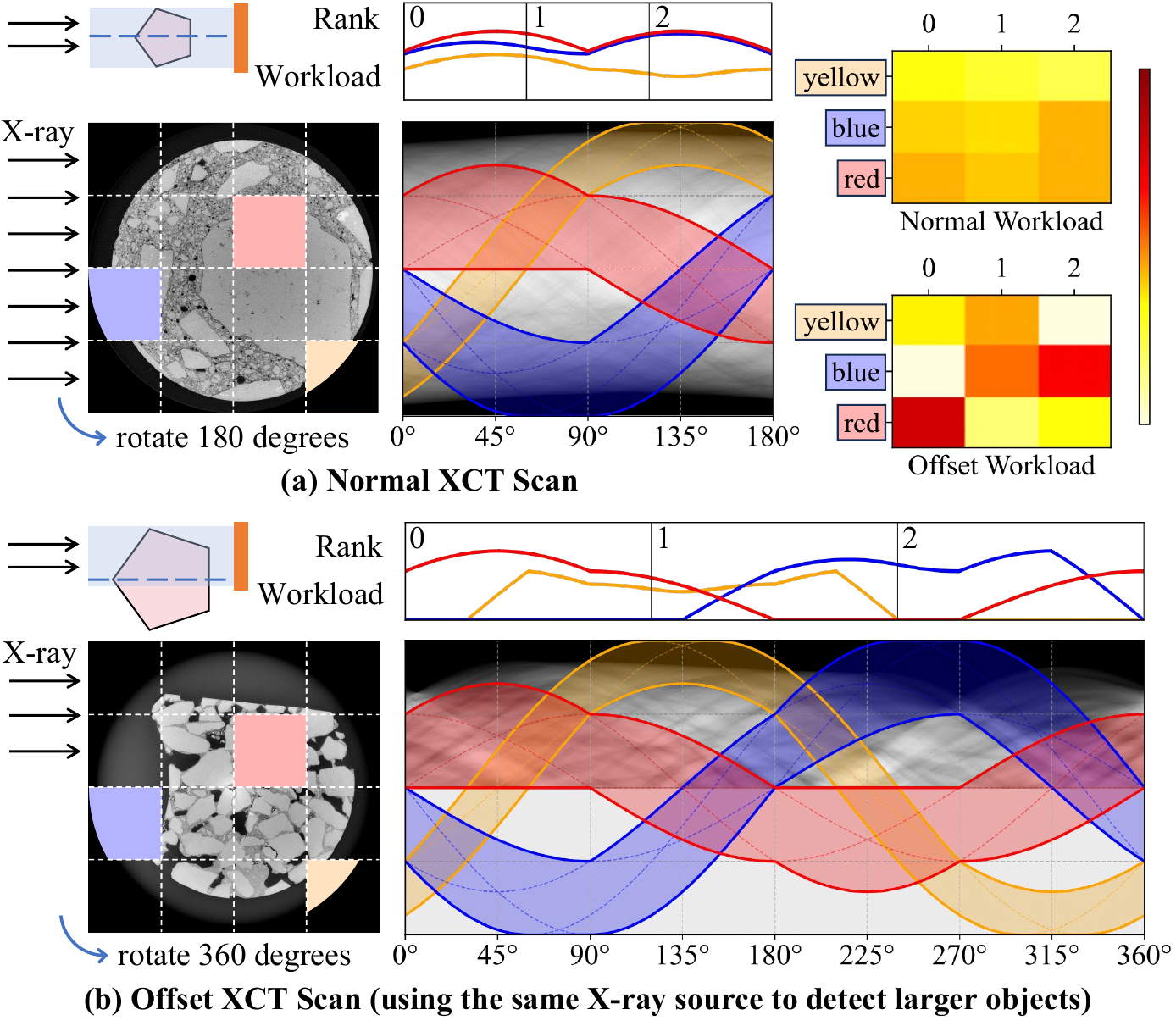}
    \caption{\small{Normal scan versus offset scan XCT. Offset scan is favorable for acquisition efficiency. Offset scan yields a more imbalanced workload in comparison to normal scan.}}
    \label{fig:workload}
  \end{center}
\end{figure}

\textbf{b) Addressing Load Imbalance in XCT Offset Scans.}\label{sec:balancing}
There are two main causes of load imbalance:
First, the increase in problem complexity leads to load imbalance. Compared with previous image reconstruction work that focused only on a single specimen, our framework handles the simultaneous reconstruction of $N_{specimen}$ specimens with various projections($N_p$), detector rows($N_r$), detector channels($N_c$), and both normal scan CT and offset scan CT. 
Second, the specimens often use offset scanning where the sample is shifted laterally (offset) to extend the field of view (FoV) beyond the detector’s physical size, hence enabling the imaging of large objects that exceed the detector’s native FoV. The downside is that offset scan yields a more imbalanced workload in comparison to normal scan. As shown in Fig.~\ref{fig:workload}, using projection parallel and slice parallel simultaneously will lead to load imbalance. The yellow, blue, and red regions represent different slice partitions, and the rank numbers at the top indicate different projection partitions. The colored areas illustrate memory access footprints, which vary in proportion to the corresponding computational workload. The heat map on the upper right shows the imbalanced workload on different MPI ranks under the combined effect of the two types of parallelism.

\underline{Group Row Partitions and Cyclic Mapping to MPI Ranks.} \label{sec:group}
To address load imbalance, we propose a group-and-cycle mapping strategy (Fig.~\ref{fig:partitioning}). First, row partitions across specimens (including normal and offset scans) are grouped to balance workloads across row-parallel MPI ranks and reduce synchronization overhead from MPI\_Barrier calls. Second, these grouped row, projection, and slice partitions are cyclically mapped to MPI ranks—ensuring ranks rotate through diverse workloads, avoiding repetition and achieving balanced computation regardless of partition counts. Third, the method improves communication efficiency by shifting from many-to-one MPI\_Reduce to all-to-all MPI\_Reduce\_scatter\_block, eliminating bottlenecks and evenly distributing workloads.

\textbf{c) Optimized Pipeline between Group Row Partitions.} \label{sec:pipeline}

Building on the group and cyclic mapping optimization (Section~\ref{sec:group}-b), computation, communication, and I/O times across group row partitions become balanced. To further boost throughput, the reconstruction process is streamlined into four stages: \underline{1) Loading Projections ($T_{\text{load}}$)}: Data is loaded once from the PFS to LLIO (node-local storage), with group row partitions evenly distributed across $P_{\text{slice}}$ nodes. \underline{2) FBP Computation ($T_{\text{comp}}$)}: Each node retrieves projections from LLIO, broadcasts them, and performs Beer-Lambert pre-processing~\cite{strong1952theoretical}, Gaussian blur, filtering~\cite{kak2001principles}, and back-projection computation $\mathcal{O}(N_p \cdot N_x \cdot N_y \cdot N_z)$, scaling across $P_{\text{proj}}$ and $P_{\text{slice}}$. \underline{3) Segmented Reduction ($T_{\text{comm}}$)}: Back-projection results from all $P_{\text{proj}}$ nodes are reduced using segmented \texttt{MPI\_Reduce\_scatter\_block}, improving efficiency over standard many-to-one communication. \underline{4) Volume Storage ($T_{\text{store}}$)}: Output slices are normalized based on CT number (HU) range~\cite{goldman2007principles}, quantized to \texttt{uint16\_t}, and written back to storage.

As shown in Fig.~\ref{fig:pipeline}b, overlapping these stages across group row partitions ensures that the overall runtime is determined by the slowest stage, maximizing throughput.

In summary, we achieved high-throughput, high-performance, high-resolution image reconstruction in \method{} framework by using group and cyclic mapping of row partitions, establishing an efficient pipeline for the simultaneous reconstruction of different specimens, including a mix of normal and offset scans. These strategies minimized the bottlenecks, optimized for load balancing, and finally achieved both high-performance and high-throughput image reconstruction (as will be shown in the evaluation section).

\begin{figure*}[t]
  \begin{center}
\includegraphics[clip,width=\textwidth]{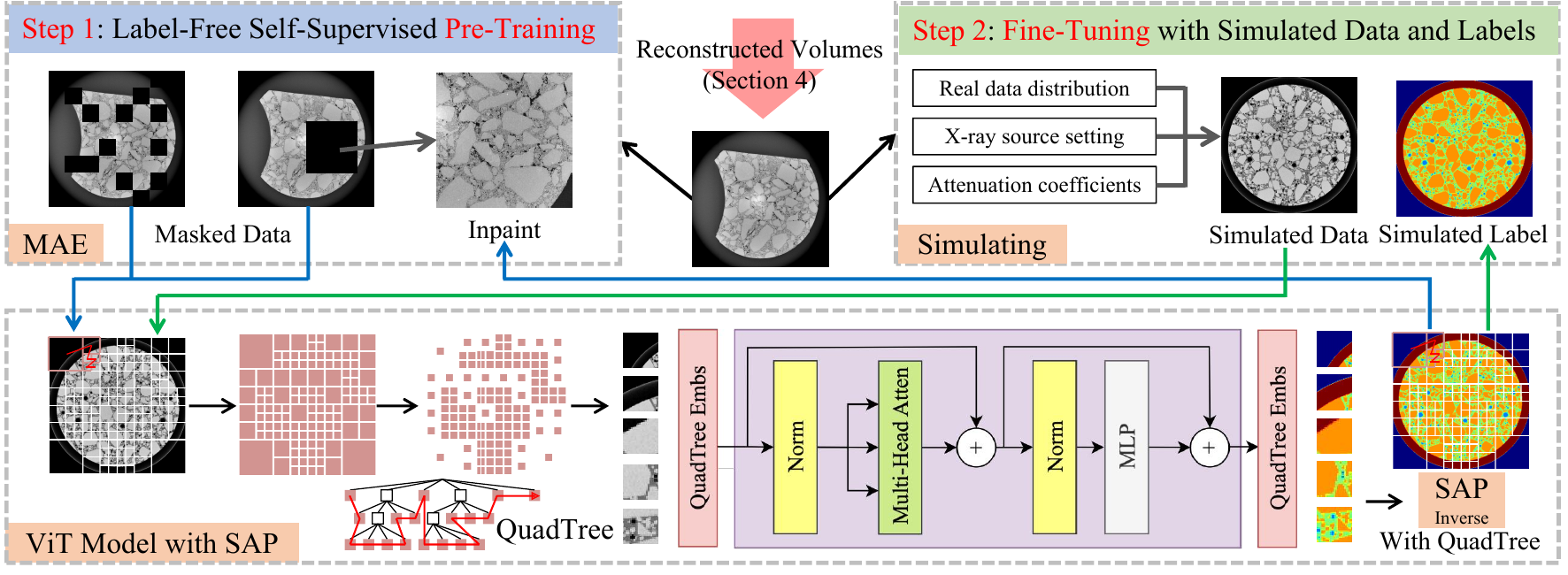}
    \caption{\small{Overview of AI foundation model training. Two steps with: 1) MAE label-free pre-training and 2) fine-tuning with simulated data.}}
    \label{fig:AI_overview}
  \end{center}
\end{figure*}

\begin{figure}[t]
  \begin{center}
\includegraphics[clip,width=0.485\textwidth]{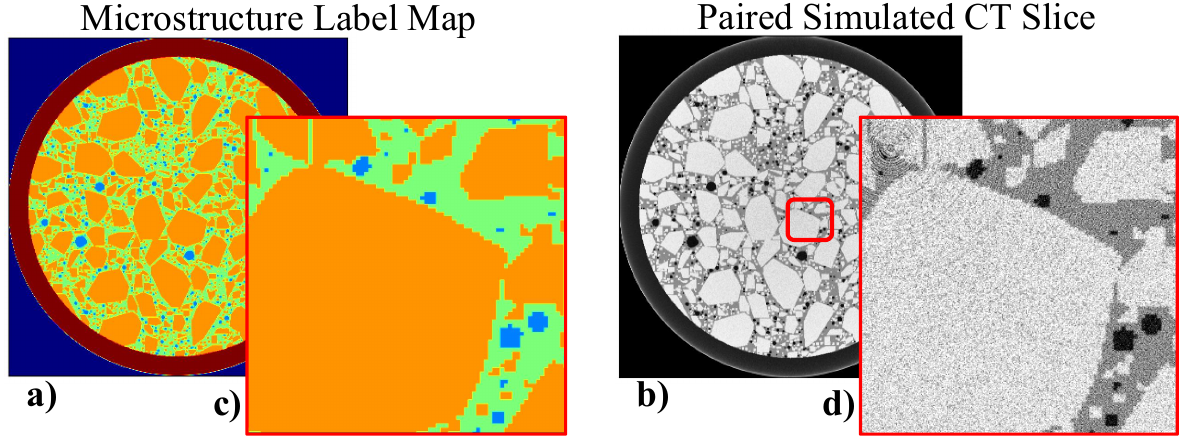}
    \caption{\small{An example slice from a simulated microstructure label map (a) and corresponding XCT slice in (b). An expanded view of the label map and CT slice in a Region Of Interest (ROI) near center shown in panels (c),(d). Noise, blurring and ring artifacts are present in the XCT (d).}}
    \label{fig:simulate}
  \end{center}
\end{figure}

\subsection{Foundation Model for High-resolution 3D Segmentation}\label{sec:AI_training}
The core requirement from an AI vision model that can enable the analytics required for robust inspection of road infrastructure is high-quality 3D segmentation. Building downstream functionalities become straightforward, once high-quality segmentation exists. High-quality segmentation is a challenging problem, for three reasons. First, high quality datasets are necessary. Second, labeling large datasets of high-resolution 3D volumes is unfeasible. Third, the memory requirements for high-resolution segmentation are notoriously restrictive. In this section we discuss the design and training process of the vision transformer foundation model we developed for this project. Note that the model, that we openly release, can be used stand-alone for segmentation of high-resolution concrete, asphalt, sand etc 3D images. 

\textbf{a) Generating Labeled Data by Simulating Reconstruction.}
\label{sec:sim}
To generate training data for the model, we simulated high-resolution synchrotron CT data of concrete by developing a scalable pipeline that transforms large-scale virtual microstructures into realistic XCT scan volumes. 
Starting from volumetric data generated using PyCMG~\cite{PyCMG} at a coarse resolution of 150 $\mu m$, the pipeline extracts multiple sub-volumes and resamples them to a resolution of 12.02 µm, matching the voxel size of real synchrotron CT scans. 
Each resampled sub-volume is randomly cropped along the z-axis to a depth between 50 and 120 slices, resulting in 3D volumes of approximately $8,192 \times 8,192$  pixels per slice. 
These high-resolution volumes are then passed through an XCT forward model using the ASTRA toolbox~\cite{van2016fast} with an FBP reconstruction algorithm, simulating realistic XCT projections and reconstructions. 
Material-specific attenuation coefficients—derived from regions of interest in real synchrotron scans of pores, cement paste, aggregates, and background—are used to assign physically meaningful intensities. 
Each volume is further processed under varying conditions, including multiple projection sparsity levels, applying Poisson and adding Gaussian noise, Gaussian blurring in the projection data, and synthetic ring artifacts to replicate common experimental degradations. 
In total, 780 unique volumes were generated, providing a diverse dataset for robust machine learning training. This comprehensive synthetic dataset serves as a valuable resource for downstream fine-tuning tasks such as segmentation as well as reconstruction, and representation learning. 

\textbf{b) Self-Supervised Learning with MAE.}
\label{sec:mae}
Training accurate segmentation models that generalize well requires large labeled datasets, but generating such data is time-consuming and biased. In our case, obtaining large labeled datasets is unfeasible, though we have access to substantial unlabeled data. To address this, we leverage unsupervised learning to pre-train models, capturing broad features for downstream tasks.
Transfer learning is a common approach in AI/ML, where features from pre-trained models are adapted to new tasks. ViTs, in particular, excel in transfer learning. We utilize pre-trained ViT models, such as SAM 2~\cite{ravi2024sam}, for segmentation. However, these models are trained on significantly different data from our high-resolution 3D datasets, prompting us to explore self-supervised pre-training with a Masked Autoencoder (MAE).

Self-supervised pre-training enables large-scale ViT training using our vast data resources, producing versatile models for various segmentation tasks, including simulated labeled data. The MAE stage of training, following the ViT-Base size, serve as proofs of concept, and scaling it further to enhance downstream segmentation performance is feasible. 

\textbf{c) Symmetric Adaptive Patching.}\label{sec:SAP}
\underline{Base Model:} We use the encoder of the well-known segmentation model SAM 2 \cite{ravi2024sam} as a base model that we adapt. 
SAM 2 offers ViT variants (ViT-Base (b), ViT-Large (l), and ViT-Huge (h)), each with 12, 24, or 32 transformer layers, respectively. These weight configurations—b, l, and h—are pre-trained using Masked Autoencoders (MAE) \cite{he2022masked}. 
\underline{Symmetrical Adaptive Patching (SAP):} SAP is a method for hierarchical image segmentation that enhances model learning by extracting multiscale spatial structures from input images. SAP takes inspiration from the use of adaptive patching in 2D pathology images~\cite{zhang2024adaptive}. The key stages are as follows.
(1) Edge Detection:
Uses Canny edge detection~\cite{canny1986computational} to extract hierarchical details as grayscale edges, smoothing out irrelevant features while preserving structural information.
(2) Adaptive Patches: AP first constructs quadtrees (2D) or octrees (3D) by recursively partitioning edge-detected images. To construct the tree $L$, AP creates tree nodes $L_n$ representing specific regions where $n$ is the number of the leaf nodes. The leaf nodes $L_{n+s-1}$ is defined recursively as follows:
\begin{equation}\begin{aligned}\tiny
L_{n+s-1} = \begin{cases} 
L_n, & \text{if $n \geq N$} \\
L_n[i] = \{L^{\text{1}}_n, L^{\text{2}}_n, ..., L^{\text{s}}_n\}, & \text{$\arg\max\limits_i V(L_n[i])$}
\end{cases}
\end{aligned}\end{equation}
where $V$ is the criterion we use to differentiate the different levels of details. We choose the maximum sum of pixel values among the tree nodes as $i = \arg\max\limits_i V(L^n[i])$.
The set $\{L^{\text{1}}_n, L^{\text{2}}_n, ..., L^{\text{s}}_n\}$ represents the $s = 2^d$ new child nodes created after subdividing $L^n[i]$, where $d$ is the number of dimensions. For example, $s = 4$ and $s = 8$ correspond to quadtree and octree structures, respectively~\cite{samet1984quadtree}.
Next, we split regions based on pixel intensity variance, ensuring uniform sequence length for GPU efficiency. Finally, we generate patch sequences via Z-order space-filling curves, maintaining alignment between image and mask patches.
(3) Depatching: We constructs masks during inference by upscaling the patch sequences using the original quad/octree structure, hence eliminating the need for U-Net decoders by using the sparse mask information instead.

\textbf{d) Training Our Foundation Model.}
In Fig.~\ref{fig:AI_overview}, we present the complete training flow of our foundation model. We trained the model in two steps. First, we perform pre-training on label-free high-resolution images using MAE (as described in Section~\ref{sec:mae}-b). The pre-training phase uses 45 sample groups, with each group trained for a total of 800 epochs. The training is conducted on 2,048 GPUs provided by Frontier. Through MAE pre-training, we enhance the model’s generalization capability on real-world data.

Next is the fine-tuning stage. We first employ the simulated reconstructed method we introduced in Section~\ref{sec:sim}-a to generate high-precision data and labels, totaling $62,400$ $8K^2$ 2D slices ($780\times80$). We then replace the encoder in SAM 2 with the pre-trained MAE weights and split the synthetic dataset into training and testing sets at a $85\!:\!15$ ratio for fine-tuning. Additionally, to address the long-sequence issue inherent in attention mechanisms, we adopt SAP to reduce computational overhead. The Evaluation section provides quantitative and qualitative comparisons between our model and baselines.

\subsection{Fusion of XCT and AI inference in to a Single Pipeline}\label{sec:end2end}
This section focuses on optimizing the image reconstruction and AI-driven analysis pipeline to eliminate redundant I/O operations. Our goal is to minimize unnecessary data transfers and enhance overall efficiency. 


\begin{figure}[t]
\begin{center}

\begin{minipage}{0.485\textwidth}
    \includegraphics[clip,width=1\textwidth]{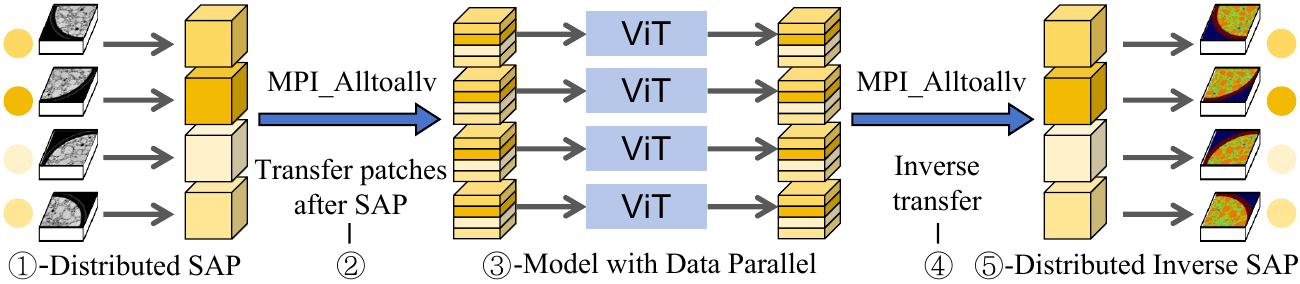}
    \caption{\small{
    Memory-resident end-to-end pipeline for distributed image reconstruction fused with 3D segmentation.
    }}
    \label{fig:merge_AI}
\end{minipage}

\vskip 5pt

\begin{minipage}{0.485\textwidth}
    \includegraphics[clip,width=1\textwidth]{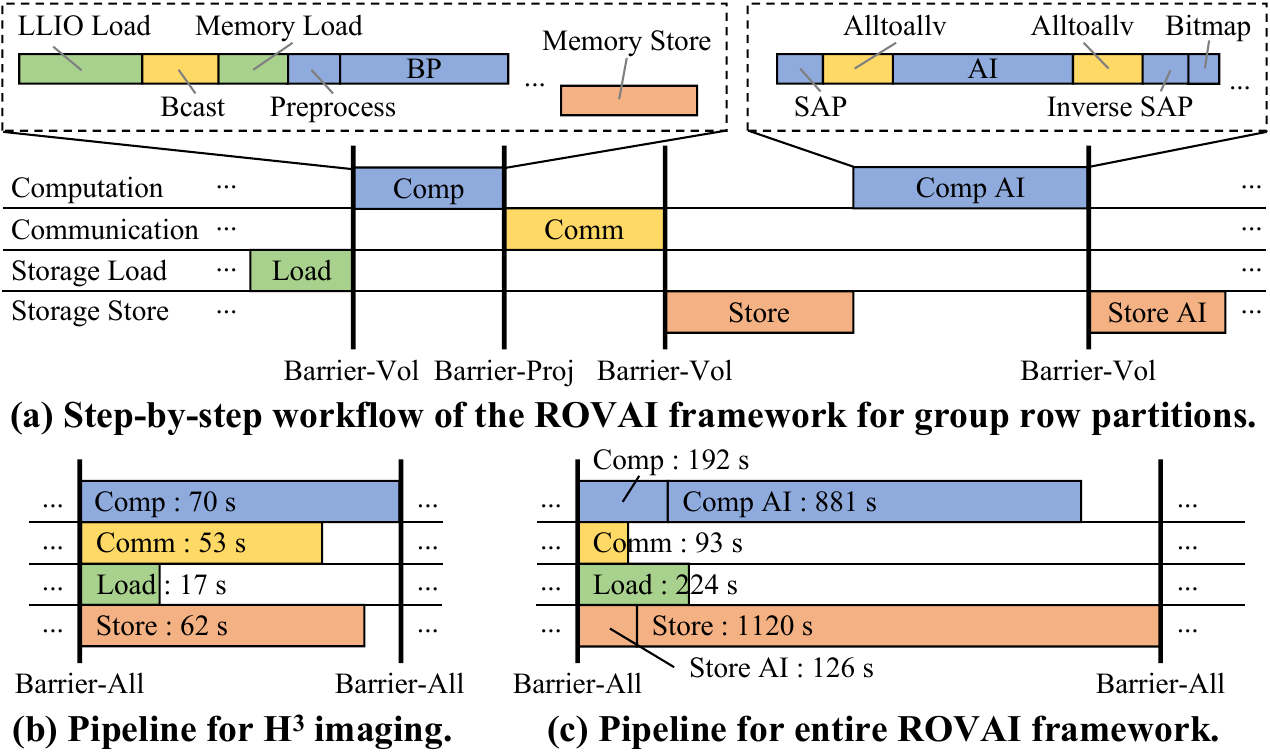}
    \caption{\small{Pipeline optimization runtime measured in seconds. (a) is the step-wise workflow aligned with different stages of our performance model. The dotted box highlights a detailed view of two computation stages. (b) Example pipeline for H\textsuperscript{3} imaging using optimal scaling strategies (P\textsubscript{proj}=48, P\textsubscript{slice}=16, $\lvert \mathbb{G} \rvert$=8) across 12,288 nodes. (c) Example pipeline for entire \method{} using optimal scaling strategies (P\textsubscript{proj}=48, P\textsubscript{slice}=4, $\lvert \mathbb{G} \rvert$=16) across full Fugaku: 152,064 nodes.}}
    \label{fig:pipeline}
\end{minipage}

\end{center}
\end{figure}



\textbf{a) Memory-resident End-to-end Pipeline.}\label{sec:workround_storage}
To mitigate performance bottlenecks caused by frequent I/O operations on the PFS, \emph{ROVAI} implements a memory-resident end-to-end pipeline, as shown in Fig.~\ref{fig:merge_AI}. In this design, high-resolution 3D volumetric data (more than 2TB per volume), generated by the FPB module, is stored entirely in the memory of the MPI ranks. This allows downstream AI tasks, like semantic segmentation, to directly access the data without having to load or store it from the PFS, eliminating redundant I/O operations.
By keeping the data in memory throughout the pipeline, \emph{ROVAI} prevents unnecessary data transfers. This is particularly important since I/O can be a significant bottleneck when performing full-system scale parallel computing for FBP image reconstruction and AI-driven 3D image analysis.

\begin{figure*}[t]
    \begin{center}
    \includegraphics[width=0.98\textwidth]{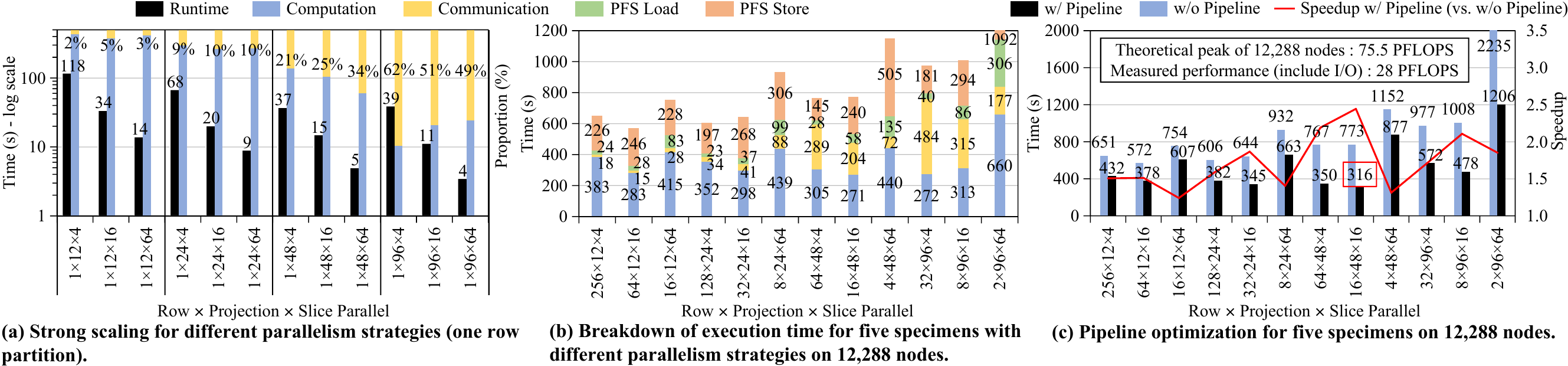} 
    \caption{\small{H\textsuperscript{3} image reconstruction performance, scaling, and effect of parallelism.}}
    \label{fig:validation}
    \end{center}
    
\end{figure*}

We implemented two key strategies for reducing storage I/O. First, to minimize redundancy in the mask-based outputs from AI models, we introduced a bitmap-based representation~\cite{ayres2002sequential}. This optimization reduces the original 32-bit output from the AI model to a 2-bit representation, corresponding to a four-class mask, thus reducing the storage footprint to $\frac{1}{16}$ of the original size. Second, once the image reconstruction results are generated, AI segmentation is performed directly in-place within memory, bypassing the need for storage access and reducing I/O overhead. The only additional storage access occurs when storing the final mask results.
These strategies result in a reduction of more than 40\% in I/O access throughout the entire \emph{ROVAI} workflow, as will be detailed in Section~\ref{sec:evaluation}.

\textbf{b) Overlapping Imaging \& AI Computations.}
The main challenge in fusing image reconstruction with AI segmentation lies in a unified scaling strategy. For model inference, we convert the trained PyTorch model to TorchScript and invoke it from C++ code. This means the model has not been modified for distributed processing. The model can only run on a single node, taking patches as input, and producing same masks of the same dimensions as output.
For image reconstruction, as we discussed in Section~\ref{sec:parallel}-a, $P_{slice}$ nodes process a single volume slice, and $P_{proj} \times P_{slice}$ nodes process an entire sub-volume. 
To align the scaling of the two components, we introduce additional communication to ensure that AI segmentation can scale effectively in alignment with the scaling strategy used for image reconstruction. Simultaneously, we minimize the communication overhead to the greatest extent possible.

To that end, as shown in Fig.~\ref{fig:merge_AI}, starting from the distributed image reconstruction results and ending with the distributed mask AI outputs, the distributed process is carried out in five steps: 
1) Distribute SAP (section \ref{sec:SAP}-c) on each $P_{slice}$ nodes. Each node retains its own quadtree structure and converts the original volume slice part into continuous patches, 2) Use \emph{MPI\_Alltoallv} communication to transfer only the continuous patches, rather than the raw volume data, across the entire $P_{slice}$ nodes. By this stage, each node has constructed its own complete continuous patches that represent slices of the volume, 3) Perform the (segmentation) model inference on each node, which is equivalent to applying data parallelism over the entire volume across $P_{proj} \times P_{slice}$ nodes, 4) Reverse transfer of the mask patches to their corresponding nodes with \emph{MPI\_Alltoallv} communication, and 5) By using the quadtree structure preserved on each node, the mask patches is transformed to the original volume 2D slice shape. 


We have detailed all stages of \method{} in Fig.~\ref{fig:pipeline}(a), we show the unified pipeline in Fig.\ref{fig:pipeline}(c). AI inference is slower than imaging due to its higher computational intensity. As a system scales to full Fugaku (152,064 nodes), \emph{ROVAI} performance is primarily bounded by I/O, and to a less extent AI inference.



\section{How Performance was Measured}

\textbf{XCT Scan Datasets from SPring-8.}
All results reported in this paper for imaging and AI training/inference are in single precision. The dataset consists of 46 specimens of highway concrete and asphalt composites, scanned using the SPring-8 BL28B2 high-energy XCT system ($\sim$200~keV). 


\textbf{Supercomputer Environment.}
Fugaku contains 158,976 A64FX nodes (ARMv8.2-A with SVE). \method{} is written in C++ based on \texttt{tcsds-1.2.41}, using the \texttt{mpiFCC-4.12.0} compiler with options \texttt{"-Nclang -Ofast -Kopenmp -Nlibomp"}. The MPI runtime is configured with the options \texttt{"--mca opal\_mt\_memcpy 1 --mca common\_tofu\_use\_memory\_pool 1"}.
Regarding the AI framework, we evaluated two configurations: (1) PyTorch v1.13 with the Fujitsu SSL2 BLAS library, and (2) PyTorch v2.6.0 with OpenBLAS 0.3.21. All inference results reported in this evaluation are based on configuration (1), PyTorch v1.13 with the Fujitsu SSL2 BLAS library.
\emph{Frontier} supercomputer, utilized for AI training, is equipped with 9,472 nodes, each housing a 64-core AMD EPYC 7453 processor, totaling 606,208 CPU cores. Additionally, the system is powered by 37,888 AMD Instinct MI250X GPUs, with each node containing 4 AMD MI250X GPUs. The AI framework is PyTorch v2.6.0 with ROCm v6.1.4.



\textbf{Performance Metrics.}
Execution time measurements are obtained using \texttt{gettimeofday()}, with timing recorded immediately after the call to \texttt{MPI\_Init()} and again after invoking \texttt{fflush(stdout)} and \texttt{MPI\_Barrier(MPI\_COMM\_WORLD)} to ensure consistent output to the PFS, and synchronization across all MPI processes.
The total number of PFS accesses is extracted from the state file generated by the Fugaku job report. Bandwidth is then computed as the total number of PFS accesses divided by the program's total execution time. The measured peak bandwidth for Fugaku's PFS I/O is taken from Fugaku report~\cite{File_System}.
Peak performance is based on Fugaku's~\cite{Fugaku} theoretical peak of 2.0~GHz and 6.144~TFLOPS per node in single precision.


\section{Performance Results}\label{sec:evaluation}

\begin{figure*}
\begin{center}

\begin{subfigure}[b]{0.37\textwidth}
    \begin{subfigure}[b]{0.48\textwidth}
        \includegraphics[width=\textwidth]{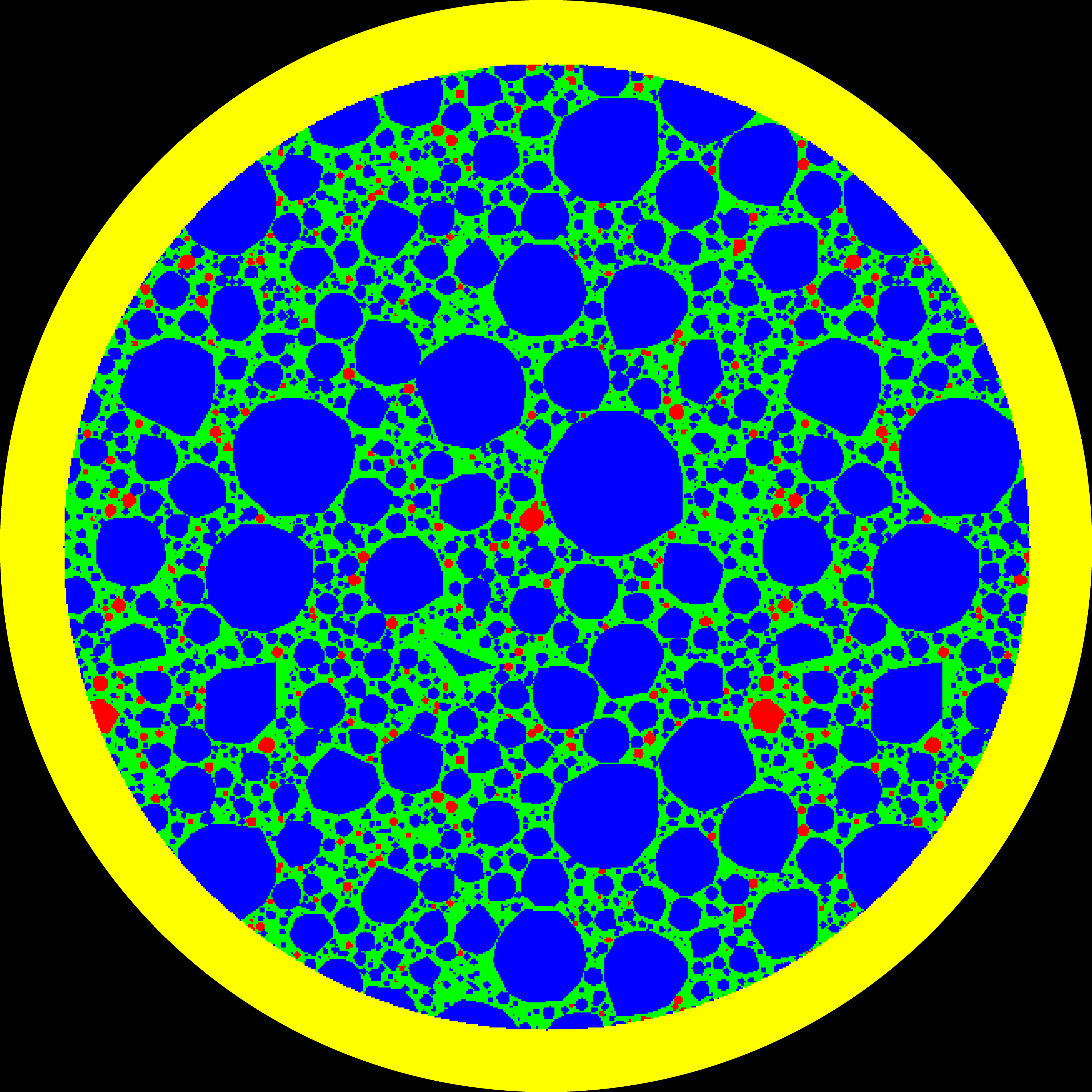}
        \caption*{\parbox[h][2\baselineskip][c]{\linewidth}{\centering \small Ground Truth.\\Dice Score: $100\%$}}
    \end{subfigure}
    \hfill
    \begin{subfigure}[b]{0.485\textwidth}
        \includegraphics[width=\textwidth]{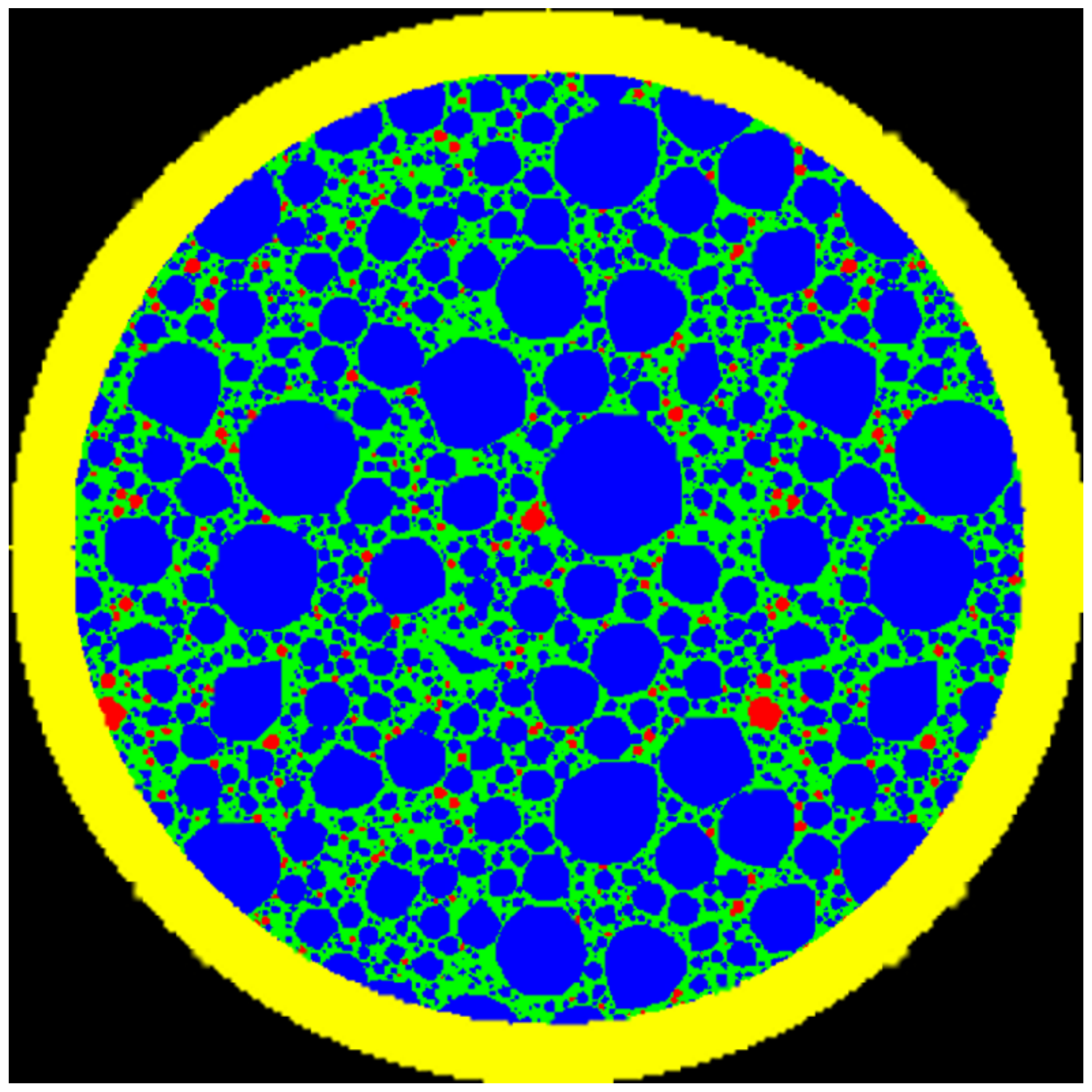}
        \caption*{\parbox[h][2\baselineskip][c]{\linewidth}{\centering \small Our model. \\Dice Score: $94.79\%$}}
    \end{subfigure}

    \begin{subfigure}[b]{0.485\textwidth}
        \includegraphics[width=\textwidth]{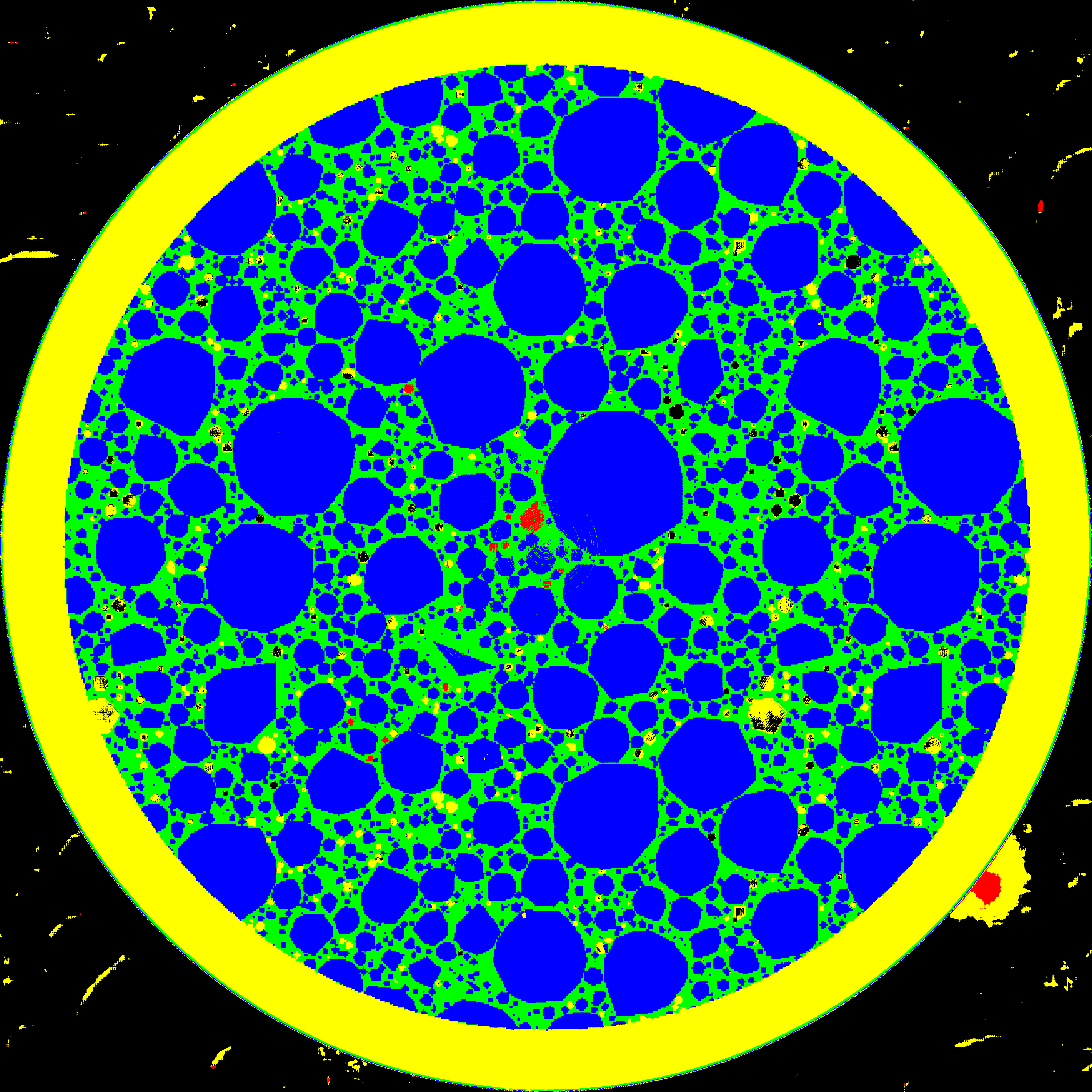}
        \caption*{\parbox[h][2\baselineskip][c]{\linewidth}{\centering \small SAM 2~\cite{ravi2024sam}. \\Dice Score: $85.98\%$}}
    \end{subfigure}
    \hfill
    \begin{subfigure}[b]{0.48\textwidth}
        \includegraphics[width=\textwidth]{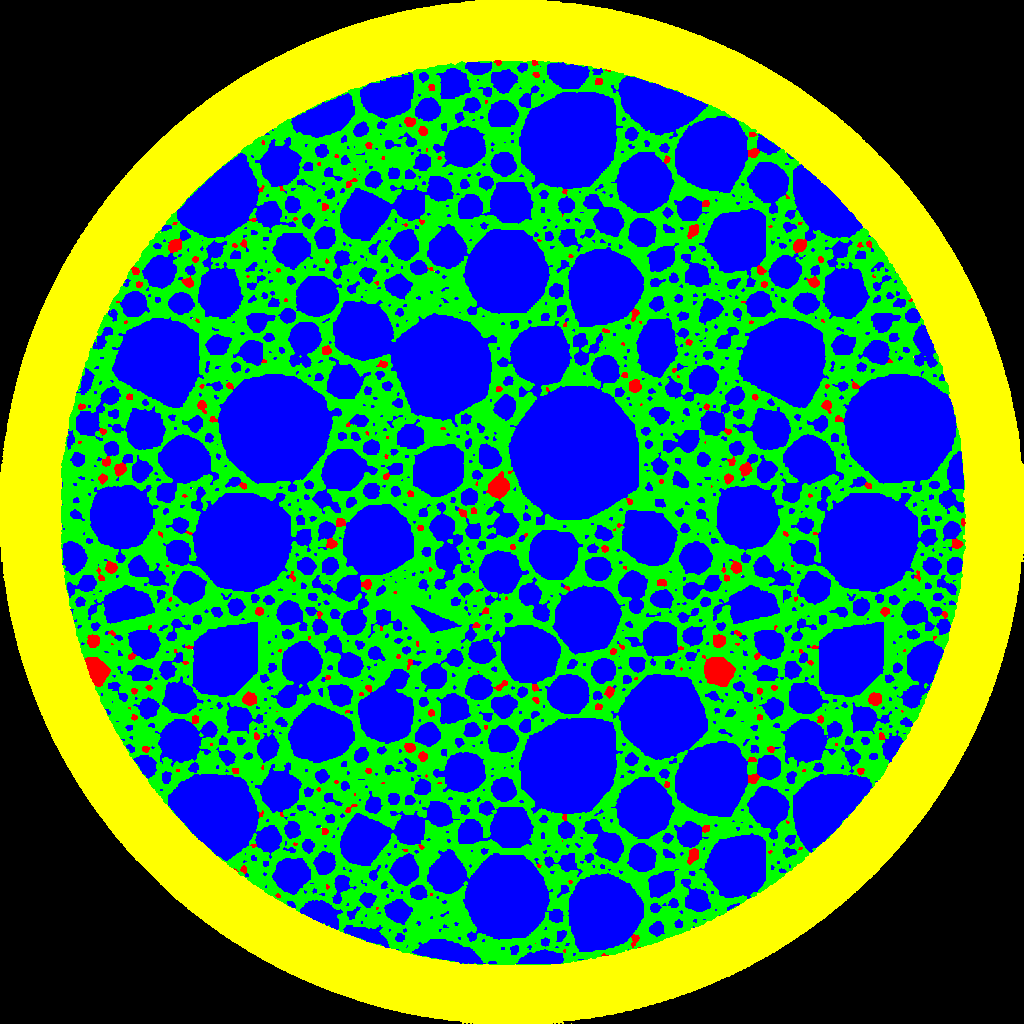}
        \caption*{\parbox[h][2\baselineskip][c]{\linewidth}{\centering \small UNet~\cite{isensee2021nnu}. \\Dice Score: $58.38\%$}}
    \end{subfigure}

    \caption{\small{Segmentation prediction on simulated samples}}
    \label{fig:seg_res} 
\end{subfigure}
\hfill
\begin{subfigure}[b]{0.61\textwidth}
    \begin{subfigure}[b]{0.49\textwidth}
        \includegraphics[width=\textwidth]{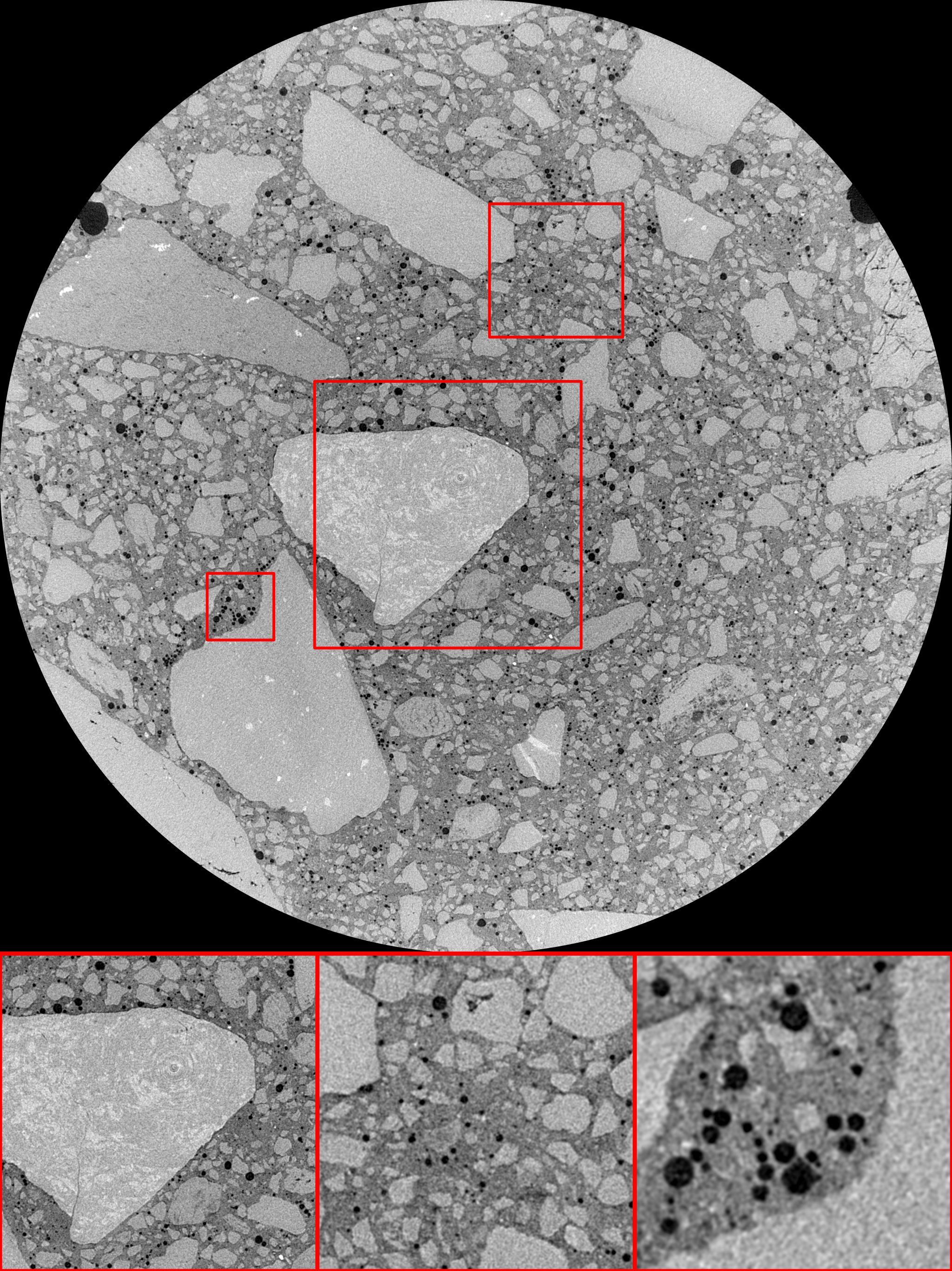}
        \caption*{\small{Real Sample Slice.}}
    \end{subfigure}
    \hfill
    \begin{subfigure}[b]{0.49\textwidth}
        \includegraphics[width=\textwidth]{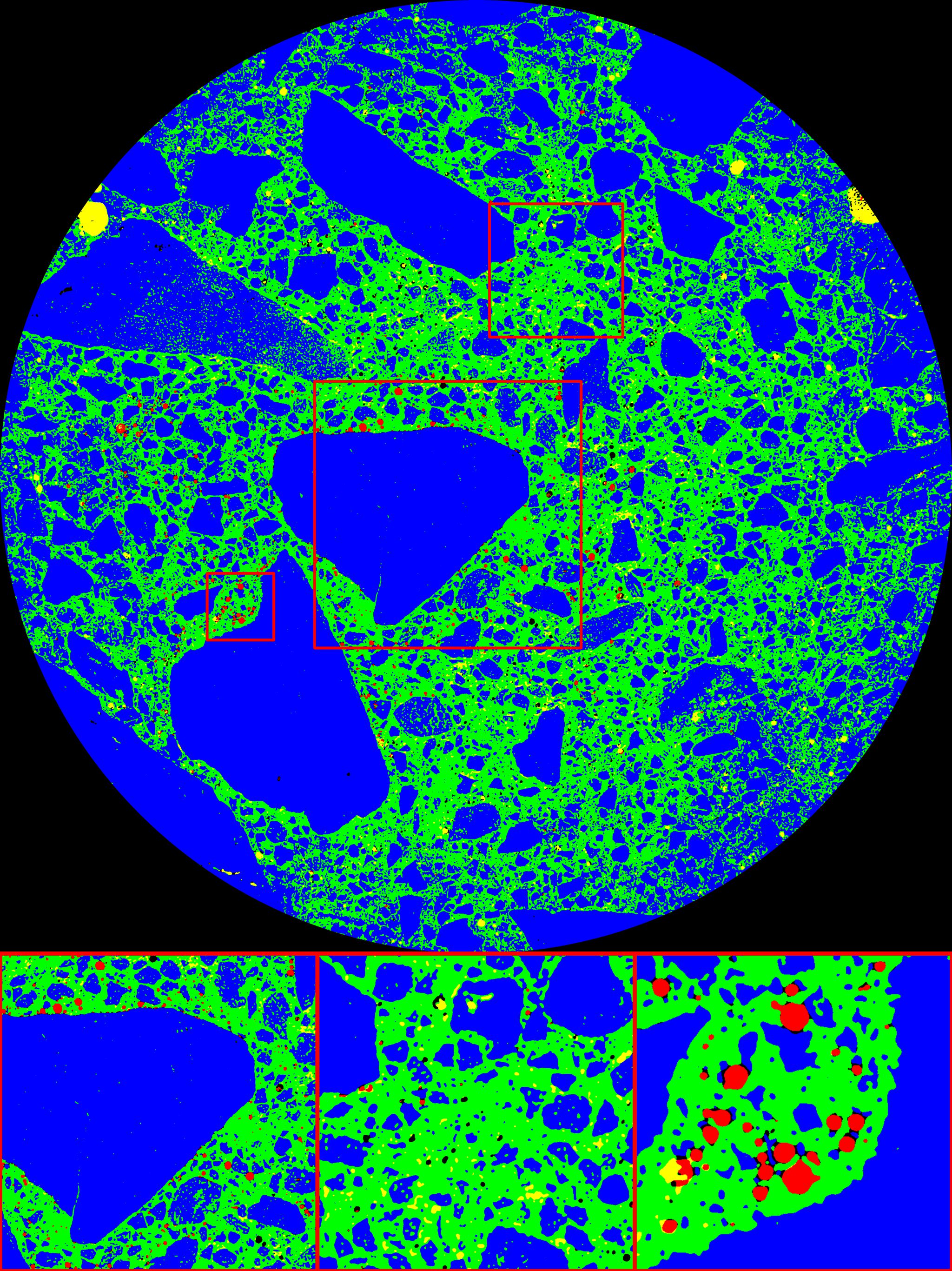}
        \caption*{\small{Zero-shot Prediction}}
    \end{subfigure}
    \caption{\small{Segmentation on real sample with zero-shot inference. The pixel-level microstructure, e.g. void area, can be precisely extracted, which is hard for human experts.}}
    \label{fig:seg_real} 
\end{subfigure}
\caption{\small{(a) Segmentation accuracy on SoTA model, SAM 2~\cite{ravi2024sam}, and our model (which uses our SAP scheme instead of the original convolution decoder). At the same GPU budget used for training, our model can go down to patch size of 2x2 (vs. 128x128 at best for SAM 2 before going OOM), for 8K resolution. As a result, our model can extract and express mask details better than SAM 2, with a big gap in accuracy favoring our model as the resolution gets higher. (b) Segmentation result on the original sample with zero-shot prediction using the model we trained on the simulated data and masks.}}

\end{center}
\end{figure*}

\begin{figure}[t]
  \begin{center}
    \includegraphics[clip,width=0.475\textwidth]{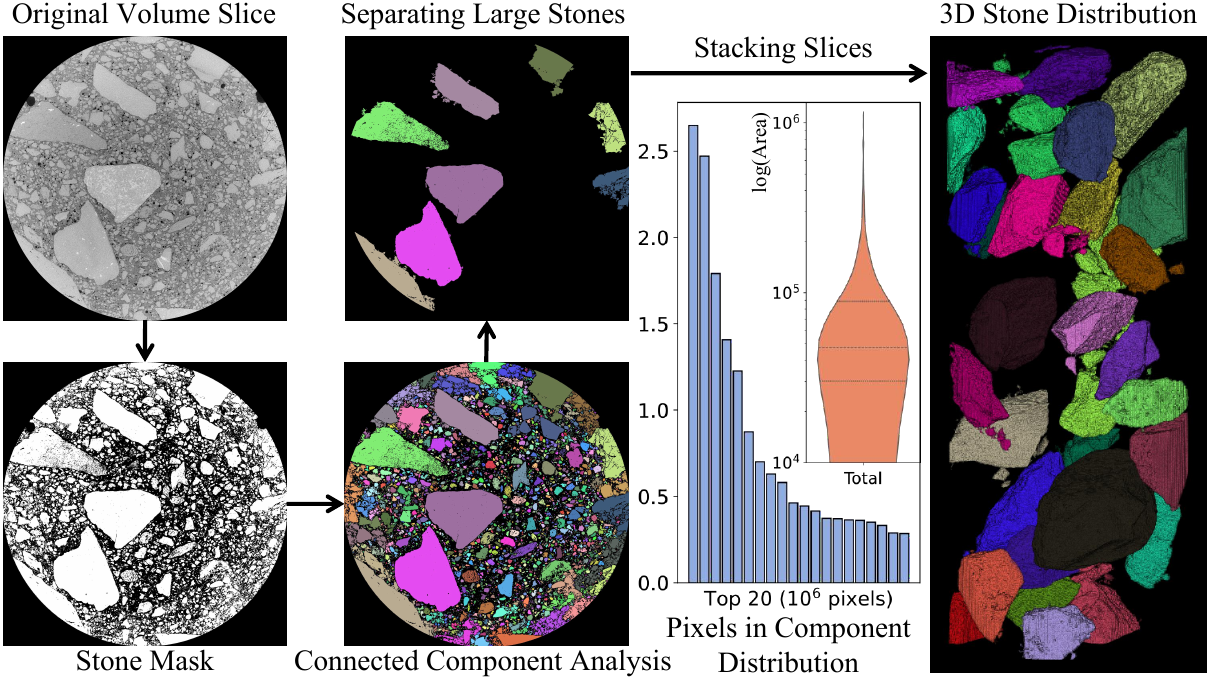}
    \caption{\small{Example of \method{}'s AI downstream task: connected component analysis~\cite{zhang2022topologypreservingsegmentationnetworkdeep}, separation, and mapping the stones distribution.
    }}
    \label{fig:downstream}
  \end{center}
\end{figure}

\begin{figure*}[t]
    \begin{subfigure}[t]{0.39\textwidth}
        \includegraphics[width=0.99\textwidth]{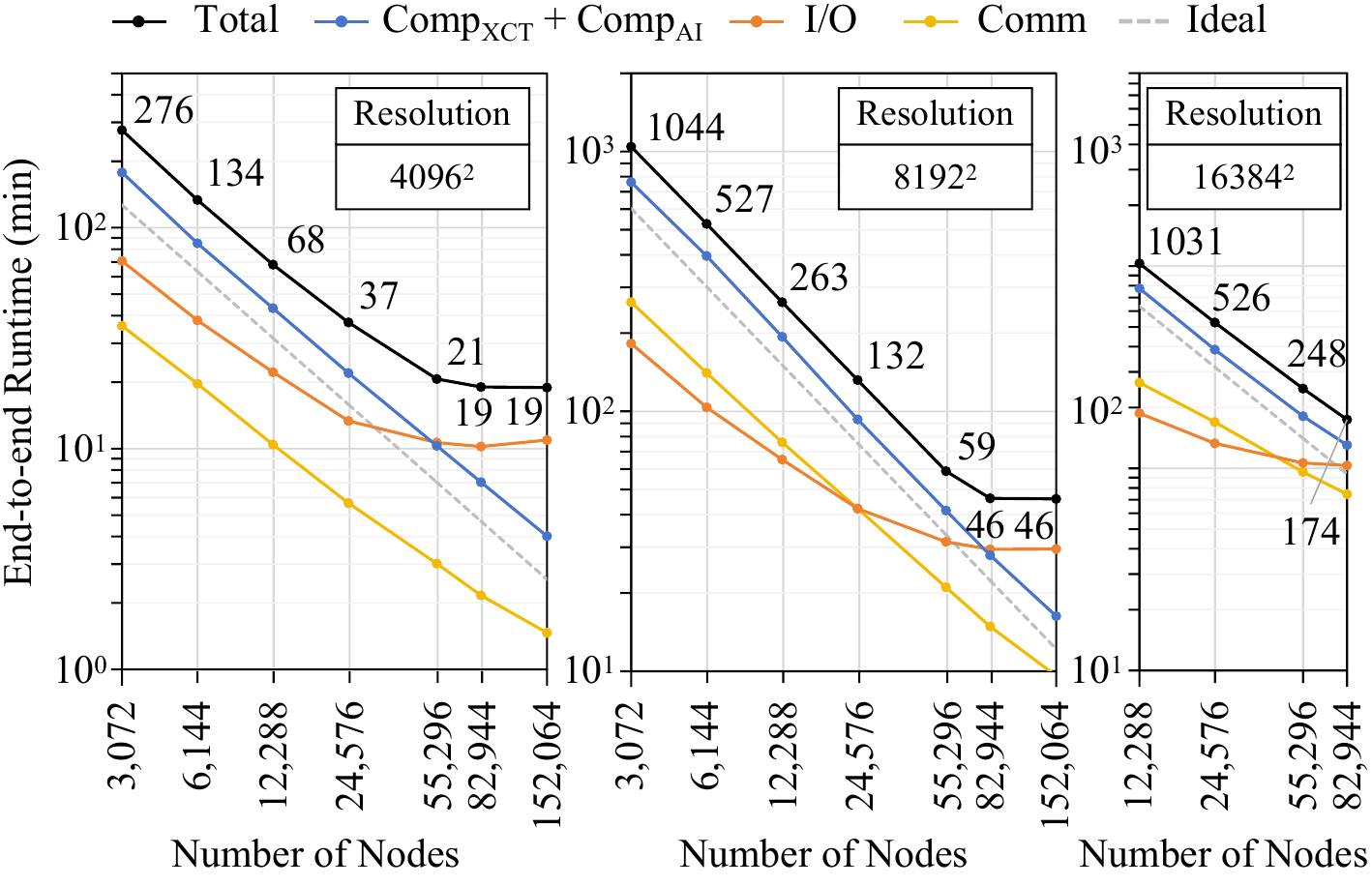}
        \caption{\small{Strong scaling evaluated by full Fugaku; \method{} \\ reconstructing 46 specimens into various resolution.}}
        \label{fig:strong_scaling}
    \end{subfigure}
    \hfill
    \begin{subfigure}[t]{0.42\textwidth}
        \includegraphics[width=0.99\textwidth]{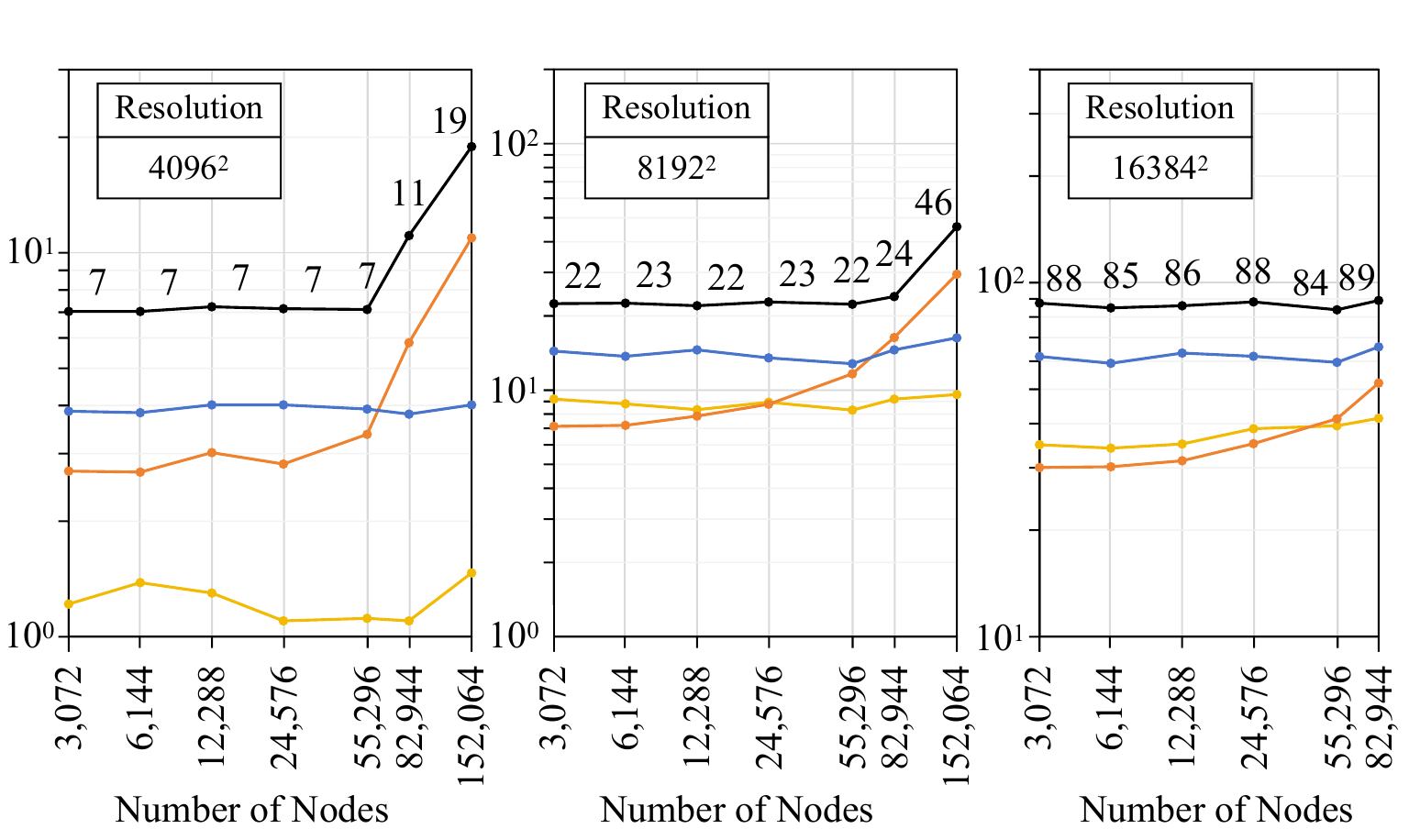}
        \caption{\small{Weak scaling for \method{}~on Fugaku. We scale by\\\hspace*{1.2em} proportionally increasing the number of rows ($\text{N}_{\text{r}}$).}}
        \label{fig:weak_scaling}
    \end{subfigure}
    \hfill
    \begin{subfigure}[t]{0.176\textwidth}
        \includegraphics[clip,width=0.98\textwidth]{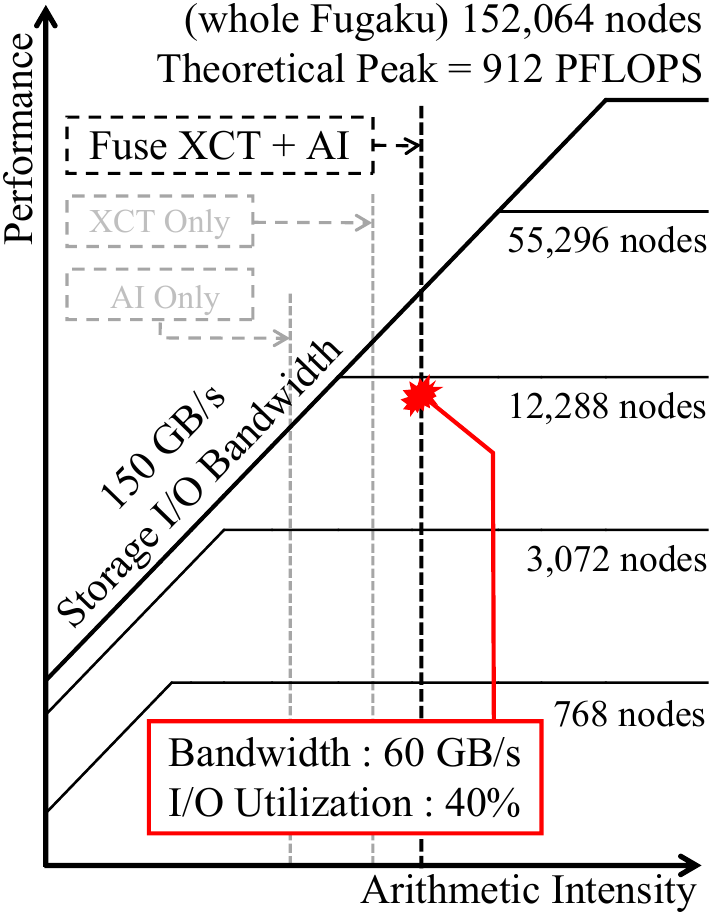}
        \caption{\small{Roofline (log-log) at $8192^2$-resolution.}}
        \label{fig:roofline}
    \end{subfigure}

    \begin{tikzpicture}[remember picture, overlay]
        \draw[dashed, thick] ($(current page.center) + (-1.87cm, +8.0cm)$) -- ++(0,4cm);
    \end{tikzpicture}

    \caption{\small{Strong and weak scaling evaluation with roofline analysis. Both fully scale up to the maximum capacity of Fugaku: 152,064-nodes.}}
    \label{fig:end2end_eval}
\end{figure*}

\begin{figure}[t]
  \begin{center}
    \includegraphics[clip,width=0.47\textwidth]{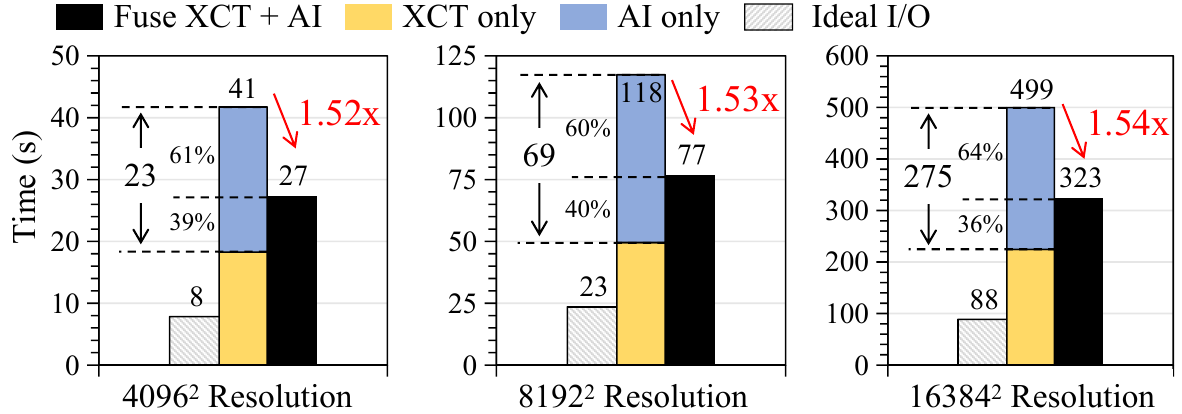}
    \caption{\small{Average runtime reduction by fusing imaging and AI inference for single specimen on Fugaku 55,296 nodes (I/O starts to dominate performance). "Ideal I/O" indicates total storage I/O access divided by the theoretical maximum storage I/O bandwidth.}}
    \label{fig:merge}
  \end{center}
\end{figure}

\subsection{Performance and Scaling of H\textsuperscript{3} FBP Imaging}
Fig.~\ref{fig:validation}a presents the strong scaling performance of the distributed FBP computation under various projection ($P_{proj}$) and slice ($P_{slice}$) partitioning strategies, along with their corresponding runtimes and communication overheads. Both partitioning approaches independently yield significant speedups, and their combined use leads to further performance enhancement. The yellow bars represent the communication overhead introduced by projection parallelism. As slice parallelism increases, both computation and communication time decrease proportionally.

Fig.~\ref{fig:validation}b and \ref{fig:validation}c further highlight the impact of these three parallelism strategies by showing the total runtime and a breakdown of the four pipeline stages described in Section~\ref{sec:pipeline}-c. 
The coordinated application of the three parallelization strategies contributes to reduced runtime across all stages of the pipeline.
In Fig.~\ref{fig:pipeline}b, if computation, communication, and PFS I/O overlap perfectly, the total runtime would be determined by the longest stage since the time spent on other stages is hidden by the ideal overlap.
However, the selection of the three parallelism parameters is constrained by node memory limits and the Message Request Queue (MRQ) size~\cite{MPIGuide}. To avoid out-of-memory and MRQ overflow errors, the group size $\lvert \mathbb{G} \rvert$ must be kept below a maximum threshold. For instance, when $P_{proj}=96$ and $P_{slice}=64$, the upper limit group size ($\lvert \mathbb{G} \rvert=16$) is still too small, weakening the effectiveness of the group and cyclic mapping optimizations described in section~\ref{sec:group}-b. This leads to a re-emergence of computation load imbalance, placing computation back in the longest stage.

Overall, the end-to-end pipeline for FBP Imaging achieves an average speedup of 1.72$\times$ (up to 2.44$\times$) across all tested combinations of the three parallelization configurations. The optimal result is obtained with the configuration $P_{\text{row}}=16$, $P_{\text{proj}}=48$, $P_{\text{slice}}=16$, and $\lvert \mathbb{G} \rvert = 8$, which reconstructs five specimens in 316 seconds, while achieving 37.1\% (28 PFLOPS) of the single precision peak performance on 12,288 nodes.
Fig.~\ref{fig:validation} confirms the effectiveness of the proposed optimization strategies outlined in section~\ref{sec:H3}.

\begin{table}[t]
\label{btcv-t}	
    	\caption{\small{Segmentation of simulated XCT dataset for multi-classes segmentation at the fine-tuning stage.}}
    \centering
    \resizebox{\linewidth}{!}{
	\begin{tabular}{|c|l|c|c|c|c|}
		\hline
        \thead{\textbf{Datset}} &
		\thead{\textbf{Model}} & \thead{\textbf{Patch Size}} & \thead{\textbf{GPU (hours)}}  & \thead{\textbf{Epochs}} & \thead{\textbf{Dice (\%)}} \\
            \cline{1-6}
            \multirow{4}{*}{\shortstack{780 unique volumes\\w/ simulated masks\\(8,192$\times$8,192$\times$(50$\sim$120))}}
            & U-Net~\cite{isensee2021nnu} & N/A & 1,280    & 500 &  58.38  \\
            \cline{2-6}
            & Swin UNETR~\cite{cao2022swin} & $256^2$ & 5,120  & 1,000 &  63.74  \\
             \cline{2-6}
            & SAM 2\cite{ravi2024sam} & $128^2$ & 5,120  & 1,000 &  85.98  \\
            \cline{2-6}
            & \textbf{Our Model} & $2^2$ & 5,120  & 1,000 &  \textbf{94.79}  \\
	       \hline
	\end{tabular}}
	\label{table:s8d_dice_score}
\end{table}

\subsection{AI Analytics}
\textbf{Segmentation Accuracy of Foundation model.}
Table~\ref{table:s8d_dice_score} compares the segmentation accuracy (Dice score~\cite{bertels2019optimizing}, ranging from 0 to 1; higher is better) of our model with representative convolution-based and ViT-based models. Training at $8\text{K}^2$ resolution often leads to out-of-memory (OOM) issues due to the large input and output sizes. To prevent OOM, convolution-based models require a reduction in both depth and channel width, which significantly degrades accuracy ($58.38\%$). ViT-based models such as SAM 2 and Swin UNTER must adopt large patch sizes (e.g. 128 or 256) to manage memory, resulting in reduced performance ($85.98\%$). Our model overcomes these limitations using the \emph{SAP} scheme, which dynamically segments the image while supporting a minimal patch size of 2. This approach alleviates the sequence length constraint in ViT models and achieves a high Dice score of $94.79\%$ at full $8\text{K}^2$ resolution. Notably, as accuracy increases, further improvement becomes more challenging, since even small gains require precise refinements along segmentation boundaries~\cite{10.1093/jmicro/dfae054}. The 9\% improvement over existing methods represents a substantial advancement, positioning our model in a qualitatively different class and enabling more sophisticated downstream analytics.

\textbf{Qualitative results.}
To further highlight the strength of our model, we present the predicted image quality in Fig.~\ref{fig:seg_res}. Although the UNet model captures the overall structure reasonably well compared to the ground truth, its limited depth and channel capacity, constrained by OOM issues, lead to poor generalization, particularly in local regions such as the "rock" structures in our task. The SAM 2 model, constrained by the quadratic complexity of the attention mechanism, is forced to use a large patch size (128) and reduce the number of upsampling convolutional layers, which also hinders its performance. In contrast, our model leverages the symmetric adaptive patching (SAP) scheme to manage sequence length effectively, achieving strong performance in both global structure and fine local detail.

Additionally, as shown in Fig.~\ref{fig:seg_real}, our model can generalize to real samples in a zero-shot setting. By performing self-regression on a large number of real samples and fine-tuning simulated labels within the simulator, the model can accurately predict microstructural masks without the need for manual annotation.



\textbf{Downstream Tasks.}
We demonstrate that by quantifying disconnected regions in statistically generated masks, our model can automate 3D object counting, separation, and distribution analysis in minutes, tasks that traditionally require weeks of expert effort at 8K resolution. In Fig.~\ref{fig:downstream}, we leverage our accurate high-resolution segmentation to extract detailed object distributions, including size, size ranking, and clustering by size. This enables precise material distribution analysis, offering valuable insights into fluid mechanics and the aging behavior of concrete materials.

\subsection{End-to-end: Imaging Fused With AI Inference}

\textbf{Performance.}
Fig.~\ref{fig:merge} presents the overall runtime, highlighting the benefits of our memory-resident fusion of imaging and AI inference. By minimizing data transfers to the PFS between the imaging and AI components, the pipeline effectively alleviates I/O contention, which is the primary bottleneck in high-resolution AI analytics of XCT imaging. This leads to more efficient utilization of I/O bandwidth and a notable improvement in overall computational performance. Specifically, the bitmap operation described in Section~\ref{sec:workround_storage}-a is applied to the segmentation mask output generated by PyTorch. This reduces redundant I/O operations. Additionally, as shown in Fig.~\ref{fig:merge_AI}c, AI computations can run concurrently with PFS storing operations from imaging. This overlap between computation and I/O stages improves throughput by reducing idle time. The framework also shows consistent runtime speedup trends across different resolutions. This stability is because I/O access scales proportionally with image resolution, and PFS I/O remains the primary bottleneck when scaling up to 55,296 nodes on Fugaku (1/3 of the system). On average, we achieve a 1.53x speedup across output images of various resolutions. When focusing solely on the AI component, 60\% of its runtime overlaps with imaging computations.

\textbf{Scaling.}
Fig.~\ref{fig:strong_scaling} shows strong scaling of \method{}~on Fugaku from 3,072 to 152,064 nodes, generating 46 volumes concurrently. For $16{,}384^3$ resolution, runs below 12,288 nodes exceeded the 24-hour limit and were excluded. Near-linear speedup is observed up to 55,296 nodes for $4{,}096^3$ and $8{,}192^3$, and up to 82,944 nodes for $16{,}384^3$, indicating efficient parallelization of XCT and AI segmentation. Performance flattens at larger scales due to PFS I/O bottlenecks that limit pipeline overlap.
Fig.~\ref{fig:weak_scaling} shows weak scaling, increasing detector rows with node count from 3,072 to 152,064. Perfect scaling is achieved at $16{,}384^3$ resolution, while lower resolutions see a drop at full scale due to insufficient workload per node, which results in a strong-scaling-like effect.

\textbf{Roofline Analysis.}
Fig.~\ref{fig:roofline} shows a full Fugaku scale Roofline. The horizontal line marks peak compute (912 PFLOPS across 152,064 nodes), while the sloped line is PFS storage peak I/O bandwidth (150 GB/s)~\cite{File_System}. Dotted markers indicate arithmetic intensity for XCT image reconstruction, AI inference, and our fused \method{} pipeline. As detailed in Section~\ref{sec:workround_storage}-a, \method{}’s optimizations shift its intensity toward the compute-I/O balance point, improving efficiency. \method{} achieves 60 GB/s throughput (40\% of peak I/O bandwidth).

\newcommand{\tbhline}{\noalign{\hrule height 1pt}}
\newcolumntype{?}{!{\vrule width 1pt}}
\renewcommand{\arraystretch}{1.15}
\newcommand{\ccmark}{$\boldsymbol{\Large \checkmark}$}
\begin{table}[t]  
 \centering
  \caption{\small{Benefits of new inspection technology over traditional methods. \( T \), \( XCT \), and \( XCT+AI \) represent traditional inspection, XCT imaging only, and XCT imaging + AI analysis, respectively.}}        
      
      \resizebox{\linewidth}{!}
      {
      \renewcommand{\arraystretch}{0.75}
      \begin{tabular}{ ?l|c|c|c? }
      \tbhline
      \bf{Road Inspection Analysis Items} & \rotatebox[origin=c]{0}{T} & \rotatebox[origin=c]{0}{XCT} & \rotatebox[origin=c]{0}{XCT+AI} \\
      \tbhline
      Detect surface degradation & \ccmark & \ccmark & \ccmark \\ 
      Visualize sub-surface conditions & & \ccmark & \ccmark \\ 
      Measure road shear strain angle & & \ccmark & \ccmark  \\ 
      Measure density of concrete and aggregate & & \ccmark & \ccmark \\  
      Measure the distribution of voids & & & \ccmark \\ 
      Measure the state of material around voids & & & \ccmark \\ 
      Measure volume ratio of aggregate, stone, etc. & & & \ccmark \\ 
      \tbhline
      \end{tabular}  
      }

\label{tab:road_inspection} 
\end{table}

\section{Implications}\label{sec:discussion}
\textbf{HPC and AI Impact.} Using full Fugaku (152,064 nodes) and SPring-8 imaging, we reconstructed 46 specimens at 8,192 resolution and ran AI segmentation in 2,762 seconds, averaging 60 seconds per specimen. For HPC, the $H^3$ imaging solution enables full-system-scale 3D reconstruction, allowing simultaneous processing of large batches of high-res XCT data for scientific analytics, and not just one image at a time as commonly done. For AI, we are the first to show that vision transformers can achieve accurate zero-shot predictions on real data when trained on simulated microstructures. Combined with adaptive patching, this demonstrates that models can learn effectively from microstructural and spatial hierarchies, offering a scalable alternative to costly pixel-level self-attention.

\textbf{Real-world Impact on Highway Inspection.} \method{} is built for ongoing use on newly collected specimens gathered regularly. Table~\ref{tab:road_inspection} lists the key analytics objectives it supports, with the Evaluation section showcasing downstream tasks that align with them. For instance, size measurement and clustering in Fig.~\ref{fig:downstream} help assess material conditions around voids and estimate volume ratios of aggregates and stones. Additional analytics results and ablation studies will be provided in the supplementary material.


\textbf{\method{}'s Outlook.} 
Infrastructure maintenance extends beyond pavements to bridges, concrete structures, and soil mechanics, all of which can benefit from \method{}’s advanced analytics, especially when guided by expert-selected downstream tasks. \method{} enables a paradigm shift in inspection, drastically reducing cost and time. It supports early crack detection, crack progression tracking, and particle size analysis for wear detection. Looking ahead, AI-driven analytics powered by \method{} could enable precise deterioration forecasting by modeling long-term environmental effects, enabling more proactive, cost-effective maintenance.

\section*{Acknowledgment}
This work was supported by JSPS KAKENHI Grant Number JP21K17750.

\bibliography{ref.bib}

\clearpage

\end{document}